\documentclass[11pt]{article}
\usepackage{amsmath,amssymb,bm,shortcuts_Social,graphicx,bigfoot,multicol,array,cite,hyperref}

\usepackage{fullpage}

\newtheorem{Lemma}{Lemma}
\newtheorem{Prop}{Proposition}
\newtheorem{Theorem}{Theorem}
\newtheorem{Def}{Definition}

\newtheorem{assumption}{H\!\!}

\makeatletter
\DeclareRobustCommand*\cal{\@fontswitch\relax\mathcal}
\makeatother

\usepackage{url}

\begin{document}
\title{RIDS: Robust Identification of Sparse Gene Regulatory Networks from Perturbation Experiments}
\author{Hoi-To Wai, Anna Scaglione,~Uzi Harush, Baruch Barzel~and Amir Leshem\thanks{H.-T. Wai and A. Scaglione are with School of ECEE, Arizona State University, USA. Emails: \texttt{\{htwai, Anna.Scaglione\}@asu.edu}. U. Harush and B. Barzel are with Department of Mathematics, Bar-Ilan University, Israel. Emails: \texttt{\{uziharush,baruchbarzel\}@gmail.com}. A. Leshem is with Faculty of Engineering, Bar-Ilan University, Israel. Email: \texttt{leshem.amir2@gmail.com}.
}}

\date{}

\maketitle              

\vspace{-1.5cm}
\begin{abstract} 
Reconstructing the causal network in a complex dynamical system 
plays a crucial role in many applications, from sub-cellular biology to 
economic systems. Here we focus on inferring gene regulation networks (GRNs) 
from perturbation or gene deletion experiments. 
Despite their scientific merit, such perturbation experiments are not often used for 
such inference due to their costly experimental procedure, 
requiring significant resources to complete the measurement of every single 
experiment.
To overcome this challenge, we develop the Robust IDentification of Sparse networks (RIDS) 
method that reconstructs the GRN from a small number of perturbation experiments. 
Our method uses the gene expression data observed in each experiment and translates 
that into a steady state condition of the system's nonlinear interaction dynamics. Applying
a sparse optimization criterion, we are able to extract the parameters of the underlying
weighted network, even from very few experiments. 
In fact, we demonstrate 
\emph{analytically} that, under certain conditions, the GRN can be perfectly
reconstructed using
$K = \Omega (d_{max})$ perturbation experiments, 
where $d_{max}$ is the maximum in-degree of the GRN, 
a small value for realistic sparse networks, indicating that RIDS can 
achieve high performance with a scalable number of experiments.
We test our method on both synthetic and experimental data extracted
from the DREAM5 network inference challenge. We show that the RIDS achieves 
superior performance compared to the state-of-the-art methods, while requiring 
as few as $\sim 60\%$ less experimental data. 
Moreover, as opposed to almost all competing methods, RIDS allows
us to infer the directionality of the GRN links, allowing us to infer empirical GRNs,
without relying on the commonly provided list of transcription factors. \vspace{-.4cm}
\end{abstract}


\section{Introduction} \vspace{-.2cm}
Our ability to understand, predict and manipulate the behavior of sub-cellular systems 
is limited by our currently partial maps of the cell's biochemical process, 
obstructing the exposure of microscopic biological mechanisms
and hindering drug development \cite{De-Smet:2010aa,buchanan2010networks,Albert:2005aa,Ideker:2008aa,Barabasi:2004aa}. 
To make advances, researchers rely on high-throughput profile technologies, such as DNA microarrays,
where a large amount of gene expression data are used for reverse engineering gene regulatory
networks (GRNs). 
The scale of these experiments allows one to infer the GRN by extracting the
correlations \cite{Marbach:2012aa} or other related statistical similarity measures 
\cite{kuffner2012inferring,friedman2000using}
in the expression levels genes under the different experimental conditions. 
Alternative methods rely on techniques such as feature selection, time series regression, etc.,
\cite{Huynh-Thu:2010aa,Haury:2012aa,Bonneau:2006aa,Yip:2010aa,singh16,Ghanbari:2015aa,
Petralia:2015aa,Wu:2016aa,Tran:2016aa,Omranian:2016aa}.  
The problem is that such methods do not directly infer the causal relationships between
genes, often falsely identifying indirect correlations as physical interactions \cite{Barzel:2013aa}.
Another challenge is that statistical similarity measures are symmetric,
and hence lack information on the directionality of the links.
Therefore, one must rely on exogenous information, such as a predefined list of 
transcription factors \cite{Marbach:2012aa,Harbison:2004aa}, 
without which it is not possible to deduce the direction of the inferred links.

To address these challenges, researchers attempt to infer the structure of the GRN directly 
from genetic perturbation data \cite{Cai:2013aa,Bonneau:2006aa,Yip:2010aa,
Gardner:2003aa,Kang:2015aa,Omranian:2016aa,Shojaie:2014aa}.
In each experiment one or more genes is perturbed, {\it e.g.}, knocked out or over/under-expressed,
and the change in the expression of all remaining genes is observed. 
Using perturbation data has two main advantages: 
(i) as opposed to correlations and other related measures,
that may result from spurious statistical dependencies, 
perturbations represent a controlled experimental measure of causal effects 
\cite{Kang:2015aa,Barzel:2009aa};
(ii) perturbations provide directional information, as  
$i$
will only respond to 
$j$
if there is a directed link/path leading from 
$j$ to $i$.
The problem is that perturbation experiments are of limited scale, with the most comprehensive
datasets incorporating at most several hundreds of perturbations, 
orders of magnitude less than the 
$\sim 10^4$
genes in a typical GRN.
Hence, despite their crucial advantages, we face severe limitations in our ability to use perturbation data
for sub-cellular network inference.


Here we exploit the fact that real GRNs are extremely sparse, with the typical gene regulated by 
at most a few other genes. 
To take advantage of this, we develop the 
Robust IDentification of Sparse networks (RIDS) method, 
that allows us to infer GRN from a small number 
of perturbations. 
Our method takes as input a set of perturbation experiments, and as output provides a minimal 
(sparse) network
$\{ {A}_{ij} \}_{i,j=1}^n$ 
that is consistent with the observed data.
Testing our approach against the DREAM5 gold standard \cite{Marbach:2012aa}
we find that with as few as
$\sim 35 \%$ of the total input provided in DREAM5, 
we are able to match, and in certain cases, even supass the top performing inference methods.
Since perturbations also encrypt information on the direction of the causal
influence relationship, RIDS allows us to 
infer the directionality of the links, as we demonstrate
by applying our method without using the DREAM5 provided list of transcription factors.   
Strikingly, we find that even without the transcription factor list, and despite using 
significantly less input data, RIDS remains competitive with the top preforming methods of DREAM5.
Finally, an additional advantage that our perturbation-based inference provides 
is that the inferred link weights $A_{ij}$ capture the true strength of the interactions,
providing an estimate for the rate constants for the different genetic reactions.
This is in contrast to the majority of inference methods that only provide likelihood estimates,
that, on their own, bare no inherent meaning, other than indicating the likelihood for the existence
or in-existence of a link. 

\noindent \textbf{Notations.} For any natural number $n \in \NN$, we denote
$[n]$ as the set $\{1,2,...,n\}$. 
Vectors (\resp matrices) are denoted by boldfaced letters (\resp capital letters). 
We denote $x_i$ as the $i$th element of the vector ${\bm x}$, 
$[ {\bm E} ]_{ {\cal S}, :}$ (\resp $[ {\bm E} ]_{ :, {\cal S}}$) 
denotes the submatrix of ${\bm E} \in \RR^{m \times n}$ 
with only the rows (\resp columns) selected from ${\cal S} \subseteq [m]$ (\resp ${\cal S} \subseteq [n]$). 
Vector ${\bf e}_{k} \in \RR^n$ is a unit vector with zeros everywhere except for the $k$th coordinate.
The superscript $(\cdot)^\top$ denotes matrix/vector transpose. $\| \cdot \|_2$ denotes the Euclidean norm
and $\| \cdot \|_1$ is the $\ell_1$-norm.\vspace{-.2cm}

\begin{figure}[!t]
\centering
\includegraphics[width=0.88\linewidth]{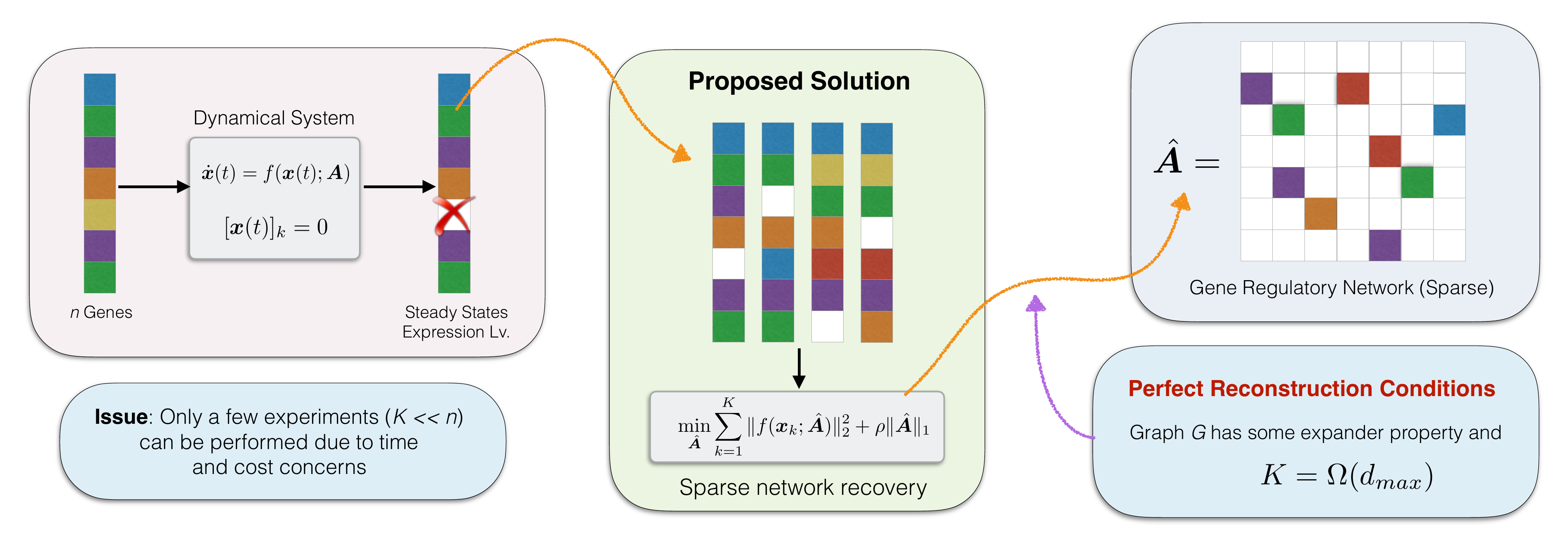}\vspace{-.3cm}
\caption{\footnotesize Conceptual overview of RIDS.
The proposed method takes the \emph{steady state} gene expression levels resulted from 
$K$ ($K \ll n$ where $n$ is the number of genes) distinct perturbation experiments and tackle
a sparse optimization problem to recover the GRN. For perfect recovery, under 
some conditions on the graph structure, the proposed solution requires only $K = \Omega(d_{max})$ experiments, where
$d_{max}$ is the maximum in-degree of the GRN.} \vspace{-.3cm}\label{fig:summary} 
\end{figure}


\section{Method \& Main Results} \label{sec:proposed}\vspace{-.2cm}
The GRN of interest is modeled as a directed graph $G = (V,E)$, 
where $V = [n] = \{1,...,n\}$ is the set of genes and $E \subseteq [n] \times [n]$ is the edge set. 
Gene $j$ is said to be a regulator (or deregulator) of gene $i$ if $(i,j) \in E$, \ie there exists an edge from
$j$ to $i$. 
The GRN is associated with a weight matrix ${\bm A} \in \RR^{n \times n}$ 
that encodes the strengths of regulation between the genes. We have
$[ {\bm A} ]_{ij} = A_{ij} = 0$ if and only if $(i,j) \notin E$. 
Here we assume the absence of self-links, hence ${\rm diag} ({\bm A}) = {\bm 0}$.
The time dependent expression level of gene $i$, $x_i(t)$, follows the nonlinear dynamic:
\beq \label{eq:mm} 
\textstyle \dot{x}_i(t) = -x_i(t) + \sum_{j=1}^n A_{ij} \cdot h( x_j(t) ; {\bm b} ) \eqs,~\forall~i \in [n] \eqs.
\eeq
where $\dot{x}_i(t) \eqdef d x_i(t) / d t$ is the rate of change of the expression level $x_i(t)$.
The first term on the right hand side captures the gene $i$'s self dynamics,
capturing processes such as degradation, and the sum captures the impact
of $i$'s interacting partners; $h(x; {\bm b})$ is the nonlinear continuous response function 
describing the regulatory mechanism such that ${\bm b}$ describes its parameters. 
For instance, setting $h(x) = c\cdot x^b$ describes
chemical activation, where according to the law of mass action $b$ is the level of 
cooperation in the activating process. Another frequently used response function
is the Hill function $h(x) = x^b / (1 + x^b)$, a saturating function such that 
$\lim_{ x \rightarrow \infty} h(x) = 1$, which captures a \emph{switch-like} activation process.

In matrix form, we can express any  \emph{unperturbed} steady state $\olx$ 
of the system \eqref{eq:mm} as:\vspace{-.1cm}
\beq \label{eq:steady}
\olx = {\bm A} {\bm h}( \olx ; {\bm b} ) \eqs,\vspace{-.1cm}
\eeq
where ${\bm h}( {\bm x}; {\bm b}) \eqdef (h(x_1; {\bm b}),~h(x_2; {\bm b}), \ldots,~h(x_n; {\bm b}))^\top$ is a column vector. 
Next, we consider a set of $K$ distinct perturbation experiments.  
In each experiment $k= 1,2,..., K$,
we \emph{fix} the state of gene $k$ 
at a desired value $z_k \in \RR$, \ie $x_{k} (t) = z_k$ for all $t$. 
The perturbed steady state ${\bm x}[k]$ takes the form
\beq \label{eq:perturb}
{\bm x}[k] = ( {\bm I} - {\bf e}_{k} {\bf e}_{k}^\top ) {\bm A} {\bm h}( {\bm x}[k] ; {\bm b} ) + z_k {\bf e}_{k} \eqs.
\eeq
Notice that when $z_k = 0$, this corresponds to the deletion of gene $k$ in the experiment. 

Suppose that the parameter ${\bm b}$ is known.
In Eq.~\eqref{eq:perturb}, the values of all ${\bm x}[k]$ are extracted from
the experimental results, hence despite the nonlinear interactions, the 
equation is linear in the unknown ${\bm A}$. 
As for each $k$, Eq.~\eqref{eq:perturb} constitutes a set of $(n-1)$-\emph{linear} 
equations in the unknown parameter ${\bm A}$.
Naively, we must conduct perturbation experiments for all $n$ genes, yielding the required
$n(n-1)$ equations to reconstruct the $n \times (n-1)$ off-diagonal 
terms of the unknown GRN. 
Such comprehensive perturbation experiments,
however, are seldom available. Indeed, the GRN of most organisms
comprises $n \sim 10^3$ genes, far exceeding the scale of 
the majority of microarray experiments, which, given the level of available 
resources, consists of $K \sim 10^1 - 10^2$ experiments.  
Hence, we shall focus on the limit where $K \ll n$ and derive a sparse
optimization algorithm to extract ${\bm A}$ from the resulting 
underdetermined linear system \eqref{eq:perturb}. 

We observe a curious property 
that can help revealing the structure of ${\bm A}$. 
Denote ${\bm a}_{col,k}$ as the $k$th column vector of ${\bm A}$.
Let $\grd {\bm h}( {\bm x}[k]) ; {\bm b} )$ be the diagonal matrix with the 
$i$th diagonal element being $h'( {x}_i[k] ; {\bm b} )$, \ie the derivative of $h( x_i[k]; {\bm b} )$ 
with respect to $x_i[k]$ evaluated at $x_i[k]$, we have:\vspace{-.4cm}
\begin{Prop} \label{prop:pert}
Consider the dynamics \eqref{eq:mm}. Assume that the perturbation in the steady states for the $k$th experiment, $\olx - {\bm x}[k]$, is small and  $\lambda_{max} ( ({\bm I} - {\bf e}_k {\bf e}_k^\top ) {\bm A} \grd {\bm h}( {\bm x}[k] ; {\bm b} ) ) < 1$. The perturbation in the steady states can be approximated by:\vspace{-.1cm}
\beq
\olx - {\bm x}[k] \approx ([\olx]_k-z_k) {\bf e}_k + ([\olx]_k-z_k) h'(z_k) {\bm a}_{col,k} \eqs.\vspace{-.1cm}
\eeq
\end{Prop}
The proof can be found in Appendix~\ref{app:pert}. 
Proposition~\ref{prop:pert} implies that the perturbation 
introduced by the $k$th experiment is \emph{limited} only to the 
\emph{direct out-neighbor} of the perturbed node $k$. 
This matches the result in \cite{baruch_universality}, 
which showed that the influence of a perturbation
on a node decays exponentially fast with respect to the 
shortest distance to the perturbed node.
In light of  Proposition~\ref{prop:pert}, 
we let $\delta > 0$ be a pre-defined threshold
and consider the index set:
\beq \label{eq:mask}  
{\cal S} = \bigcup_{j =1}^K \Big\{ (i,j) \in [n] \times [n] : \frac{\big[ \olx - {\bm x} [j] \big]_i}{ [\olx]_{j} - z_j } < \delta \Big\} \eqs.
\eeq
We also define ${\cal S}_i$ as the restriction of ${\cal S}$ to the $i$th row of ${\bm A}$, notice
that ${\cal S}_i \subseteq [K]$. 
Importantly, ${\cal S}$ can be treated as an estimate 
for the locations of \emph{zeros/non-edges} in the GRN ${\bm A}$. 

Let
$( \olx, \{ {\bm x}[k] \}_{k=1}^K )$ be the \emph{response matrix}
in which we gather 
the gene expression data 
from $K$ perturbation experiments. 
When ${\bm b}$ is known, we can recover ${\bm A}$ by solving for each $i \in [n]$:
\begin{subequations}  \label{eq:sense}
\begin{align}
\ds \min_{ \hat{\bm a}_i } \ds~~\| \hat{\bm a}_i \|_1 ~~{\rm s.t.} & ~~[\olx]_i = \hat{\bm a}_i^\top {\bm h}( \olx ; {\bm b} ),~({\bm x}[k])_i = \hat{\bm a}_i^\top {\bm h}( {\bm x}[k] ; {\bm b} ),~\forall~k \in [K] \setminus \{i\}  \label{eq:sense_a} \\
& ~~[\hat{\bm a}_i]_i = {0},~[\hat{\bm a}_i]_{j} = 0,~\forall~j \in {\cal S}_i,~\hat{\bm a}_i \in \RR^{n} \eqs. \label{eq:sense_b} 
\end{align}
\end{subequations}
The solution to the above problem, $\hat{\bm a}_i$, serves as an estimate for the $i$th row of ${\bm A}$, \ie ${\bm a}_i$. 
We emphasize that the sparsity pattern of the optimum $\hat{\bm a}_i$, 
$\{ j \in [n] : [\hat{\bm a}_i]_j \neq 0 \}$, can be interpreted as 
the set of regulators/deregulators to gene $i$ and the ranked list of non-zero
elements of  
$\hat{\bm a}_i$ indicates the probability of an existing edge in the GRN. 
The problem can be solved efficiently in parallel 
using general software package such as 
\texttt{cvx} (Available at \texttt{http://www.cvxr.com/cvx}).   

\subsection{A sufficient condition for \emph{perfect} GRN Recovery} \label{sec:theory}
To understand the fundamental limits of recovering the GRN with \eqref{eq:sense},
we study the scenario when the steady states $\olx$ and ${\bm x}[k]$ are obtained \emph{with no
noise}, \ie they satisfy the equalities \eqref{eq:steady} and \eqref{eq:perturb}.
The challenge in the analysis is that the \emph{undetermined} linear system \eqref{eq:sense_a} 
depends on the true network ${\bm A}$ itself which is a sparse matrix, and the
dynamical system is non-linear. 
We develop a new sparse recoverability condition given
that ${\bm b}$ is known. 
To describe the result, let us state the simplifying assumptions below. 
\begin{assumption} \label{ass:simp}
(a) Set ${\cal S}$ is the complement of the support of ${\bm A}$; 
(b) matrix ${\bm A}$ 
is non-negative;
(c) the approximation in Proposition~\ref{prop:pert} is exact; (d) ${\bm h} ( {\bm x} ; {\bm b} )$ admits
an exact first order Taylor approximation; (e) $\overline{x}_k - z_k \geq 0$ for all $k \in [K]$.\vspace{-.4cm}
\end{assumption}
\begin{Theorem} \label{thm:id}
Assume H\ref{ass:simp}. 
For each $i \in [n]$, if
the induced bipartite graph $G( {\cal S}_i, [n] )$ is an 
$(\alpha, \delta)$-unbalanced expander graph with left degree 
bounded in $[d_l, d_u]$ such that \vspace{-.1cm}
\beq \label{eq:id_cond}
(1+ (d_l / d_u) \delta) \| {\bm a}_i \|_0 \leq \alpha n,~~2 (d_l / d_u) \delta > \sqrt{5} - 1 \eqs, \vspace{-.0cm}
\eeq
then solving \eqref{eq:sense} with the additional constraint $\hat{\bm a}_i \geq 0$
yields a unique solution that ${\bm a}_i = \hat{\bm a}_i$.\vspace{-.3cm}
\end{Theorem}

The formal definition of an expander bipartite graph can be found in \cite{Gilbert2010,tsipn_submit} 
or in Definition~\ref{def:exp}
of the appendix. 
The proof of Theorem~\ref{thm:id} is relegated to Appendix~\ref{app:id}. 
A curious fact about Theorem~\ref{thm:id} is that the condition \eqref{eq:id_cond} depends 
on the graph structure of ${\bm A}$ as well as the sparsity of each row ${\bm a}_i$ of ${\bm A}$.
This is due to the fact that the linear system \eqref{eq:sense_a} depend on ${\bm A}$ itself. 

A possible scenario satisfying the conditions in Theorem~\ref{thm:id} 
requires choosing the set of perturbed genes $[K]$
such that each gene in $V$ is regulated by a similar number ($\sim d$) of the genes in $[K]$. 
As we shall argue in Appendix~\ref{app:id_insight}, fix $\alpha \in (0,1)$, 
when $|{\cal S}_i| \geq c(d) \cdot \alpha n$ for some $c(d) > 1$
then $G( {\cal S}_i, V )$
is an $(\alpha, 1-1/(d-1))$-expander with $d_l = d-1$, $d_u = d$ in high probability. 
Applying Theorem~\ref{thm:id} shows that perfect recovery is possible if 
\beq \label{eq:highlv}
K \approx |{\cal S}_i| > c(d) \cdot (1 + \delta ( d -1 ) / d ) \| {\bm a}_i \|_0 \geq  c(d) \cdot ( (2d-2 )/ d)\| {\bm a}_i \|_0,~\forall~i \in [n] \eqs,
\eeq 
where $\| {\bm a}_i \|_0$ counts the number of non-zeros in the vector ${\bm a}_i$. 
For example, when $d=10$, $\| {\bm a}_i \|_0 = 0.1 n$, we have 
$c(d) \cdot ( (2d-2 )/ d) \approx 2.91$. 
Since $\| {\bm a}_i \|_0 \leq d_{max}$, where
$d_{max}$ is the maximum in-degree, it shows
that $K = \Omega( d_{max} )$ is a sufficient condition for perfect recovery. 

Finally, we remark that the results in Theorem~\ref{thm:id} gives a condition
for the perfect recovery or identification of ${\bm A}$. 
This is significantly more challenging than merely \emph{reconstructing}
the GRN as required by DREAM5 \cite{Marbach:2012aa}, 
which focuses only on detecting the edges $E$ of the GRN. \vspace{-.1cm}

\begin{figure}[!t]
\centering
\includegraphics[width=0.85\linewidth]{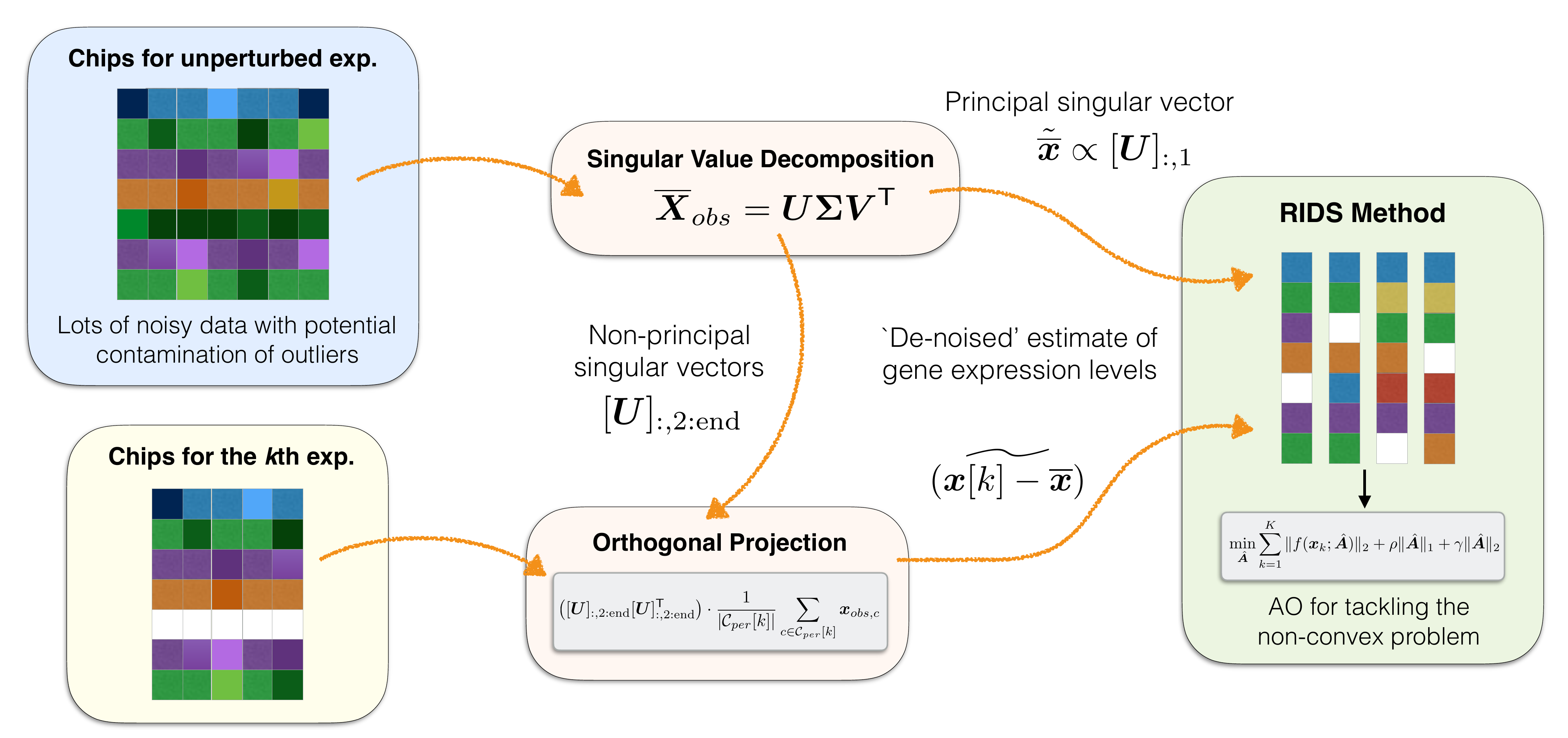}\vspace{-.2cm}
\caption{\footnotesize
Overview of the proposed RIDS method applied on empirical data. 
In the pre-processing stage, we first recover
the unperturbed steady-state expression levels $\tilde{\olx}$ by analyzing the principal component 
of the stacked response matrix $\overline{\bm X}_{obs}$. 
Then, an orthogonal projection is applied to the perturbed steady-state expression levels to recover ${\bm x}[k]-\olx$ for each perturbation condition. This forms $(K+1)$ vectors where each of them correspond to a distinct perturbation condition (including no perturbation). Finally, we tackle the robust sparse network recovery problem \eqref{eq:sense_r} for GRN recovery via the AO procedure \eqref{eq:ao}.} \vspace{-.2cm}\label{fig:dream5}
\end{figure}

\subsection{Robust IDentification of Sparse networks (RIDS) method} \label{sec:robust}\vspace{-.2cm}
So far, the theoretical model above assumes that the steady state expression data are measured 
noiselessly and the parameter ${\bm b}$ in the model response function $h(x; {\bm b})$ 
is known. 
In practice, ${\bm b}$ is unknown and maybe different for every gene. 
We have the \emph{noisy} expression data:\vspace{-.1cm}
\beq \label{eq:measure}
\tilde{\olx} = \olx + \overline{\bm{\epsilon}}~~~\text{and}~~~\tilde{\bm x}[k] = {\bm x}[k] + \bm{\epsilon}[k] \eqs,\vspace{-.1cm}
\eeq
where the vectors $\overline{\bm{\epsilon}}$, $\bm{\epsilon}[k] $ represent some additive noise of unknown distribution. 
Let $i \in [n]$, ${\bm b}_i$ be the parameter of gene $i$ and define the matrix/vector:
\beq
{\bm y}_i \eqdef \left( 
\begin{array}{c}  
 \tilde{x}_i[1]  \\ 
\vdots \\  
\tilde{x}_i[K] \\
\tilde{\overline{x}}_i 
\end{array} 
\right)~~\text{and}~~
{\bm H}_i ( {\bm b}_i) \eqdef \left(
\begin{array}{c} 
{\bm h}( \tilde{{\bm x}}[1] ; {\bm b}_i )^\top  \\ 
\vdots \\  
{\bm h}( \tilde{{\bm x}}[K] ; {\bm b}_i )^\top \\
{\bm h}( \tilde{\olx} ; {\bm b}_i )^\top 
\end{array} 
\right) \eqs.
\eeq 
Let $\hat{\bm b}_i$ be an estimate of ${\bm b}_i$. 
Naturally, one would like to relax the equalities in \eqref{eq:sense_a} and minimize
$\lambda \| \hat{\bm a}_i \|_1 + \| {\bm y}_i - {\bm H}_i (\hat{\bm b}_i) \hat{\bm a}_i \|_2 \eqs$. 
However, we observe that the $k$th element of ${\bm y}_i$ is expressed as: \vspace{-.1cm}
\beq \label{eq:obs_model}
\tilde{x}_i[k] = {\bm h}( {\bm x}[k] ; {\bm b}_i )^\top {\bm a}_i + \epsilon_i[k]
= {\bm h}( \tilde{\bm x}[k] ; \hat{\bm b}_i )^\top {\bm a}_i + \bm{\delta}[k]^\top {\bm a}_i +  \epsilon_i[k] \eqs,\vspace{-.1cm}
\eeq
where $ \bm{\delta}[k] \eqdef {\bm h}( {\bm x}[k] ; {\bm b}_i ) - {\bm h}( \tilde{\bm x}[k] ; \hat{\bm b}_i ) $ is an unknown vector 
that scales with the magnitude of $\bm{\epsilon}[k]$. 
The difference vector ${\bm y}_i - {\bm H}_i ({\bm b}_i) \hat{\bm a}_i $ is dependent on ${\bm a}_i$ 
and can not be modeled as an additive noise. 

On the other hand, the unknown parameter ${\bm b}_i$ lies 
in a parameter set ${\cal B}$. 
As such, our robust identification of sparse networks (RIDS) method 
tackles ---
for each $i \in [n]$: \vspace{-.1cm}
 \begin{subequations}  \label{eq:sense_r}
\begin{align}
\ds \min_{ \hat{\bm a}_i , \hat{\bm b}_i } & \ds~~J( \hat{\bm a}_i ; \hat{\bm b}_i ) \eqdef \| {\bm y}_i - {\bm H}_i (\hat{\bm b}_i) \hat{\bm a}_i \|_2 + \rho \| \hat{\bm a}_i \|_1 + \gamma \| \hat{\bm a}_i \|_2 \\
{\rm s.t.} & ~~[\hat{\bm a}_i]_i = {0},~[\hat{\bm a}_i]_{j} = 0,~\forall~j \in {\cal S}_i,~\hat{\bm a}_i \in \RR^{n},~\hat{\bm b}_i \in {\cal B} \eqs,\label{eq:sense_r_b} \vspace{-.1cm}
\end{align}
\end{subequations}
where $\rho, \gamma > 0$ are fixed regularization parameter. 
The derivation details of \eqref{eq:sense_r} can be found in Appendix~\ref{sec:derive}. 
This formulation is akin to the matrix uncertainty (MU) selector 
in \cite{mathieu10}
for sparse recovery with uncertainty. 
Despite being robust to measurement error, the above formulation 
simultaneously solves for the best model parameter that fits with the expression data. 

However, \eqref{eq:sense_r} is a \emph{non-convex} problem 
due to the multiplicative coupling in the least square objective function. The problem cannot be solved
directly using off-the-shelf packages. The RIDS method
applies an alternating optimization (AO) approach to get
around with the issue, \ie by running the following procedure for each $i \in [n]$ --- let $L$ be the maximum number of iterations:
\begin{center}
\fbox{
%
\begin{minipage}{.8\linewidth}\vspace{-0.2cm}
\beq \label{eq:ao} 
\begin{split}
{\sf for}~~ & \ell = 1,2,3, \ldots, L \\
& \textstyle \hat{\bm a}_i^{\ell+1} \leftarrow \arg \min_{ \hat{\bm a}_i }~J( \hat{\bm a}_i ; \hat{\bm b}_i^\ell )~~{\rm s.t.}~~\eqref{eq:sense_r_b}~{\sf satisfied} \eqs, \\
& \textstyle \hat{\bm b}_i^{\ell+1} \leftarrow \arg \min_{ {\bm b} }~ \| (\hat{\bm b}_i^\ell - \epsilon \cdot \grd_{\bm b} J( \hat{\bm a}_i^{\ell+1}; {\bm b}^\ell ) ) - {\bm b} \|_2~~{\rm s.t.}~~{\bm b} \in {\cal B} \eqs,
\end{split}
\eeq \vspace{-.1cm}
\end{minipage}
}
\end{center}
where $\epsilon > 0$ is a fixed step size and $\grd_{\bm b} J( \hat{\bm a}_i^{\ell+1}; \hat{\bm b}_i^\ell ) $
is the gradient of the cost function. 
The last step is a \emph{projected gradient} update step for $\hat{\bm b}_i$. 
The RIDS method is summarized in Figure~\ref{fig:dream5}.
In the first stage, we apply 
a pre-process method to de-noise the experimental data
(see Appendix~\ref{sec:dream5}); 
in the second stage, we tackle the robust GRN recovery problem \eqref{eq:sense_r} using the AO procedure
in \eqref{eq:ao}.

\vspace{-.2cm}
\section{Numerical Results \& Discussions}\label{sec:num} \vspace{-.2cm}
\begin{figure}[!t]
\centering
\includegraphics[width=0.45\linewidth]{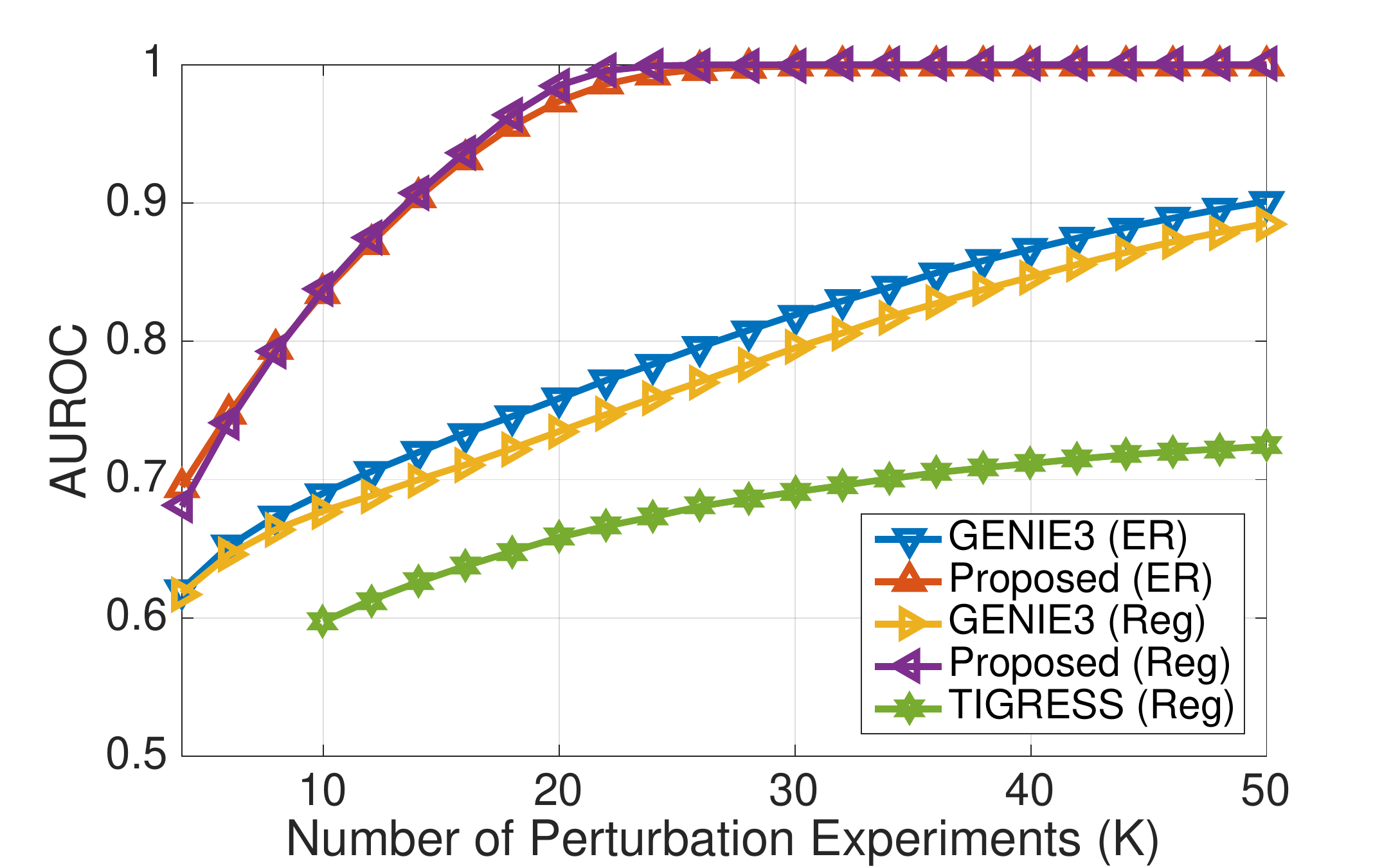}~
\includegraphics[width=0.45\linewidth]{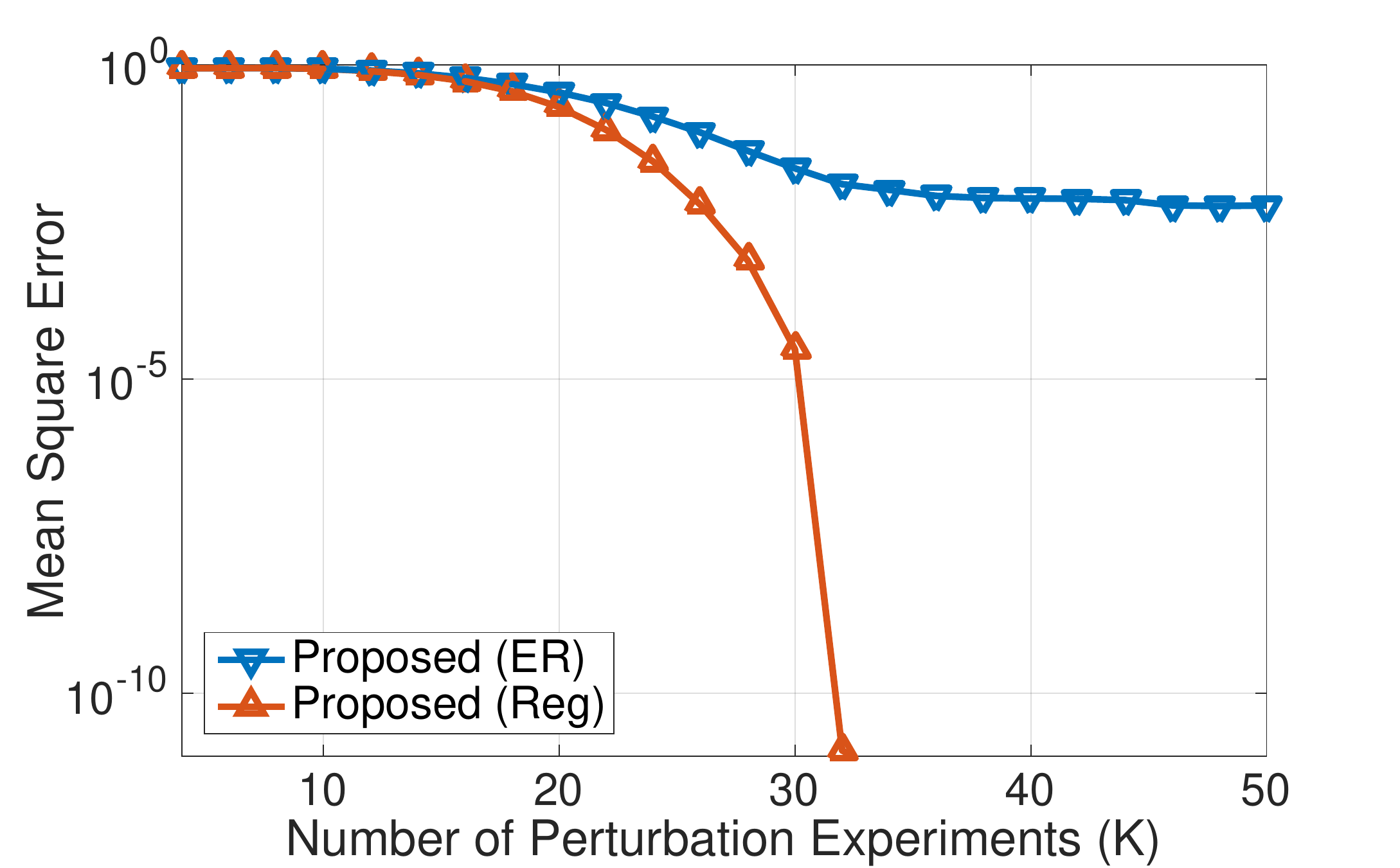}\vspace{.1cm}
\includegraphics[width=0.45\linewidth]{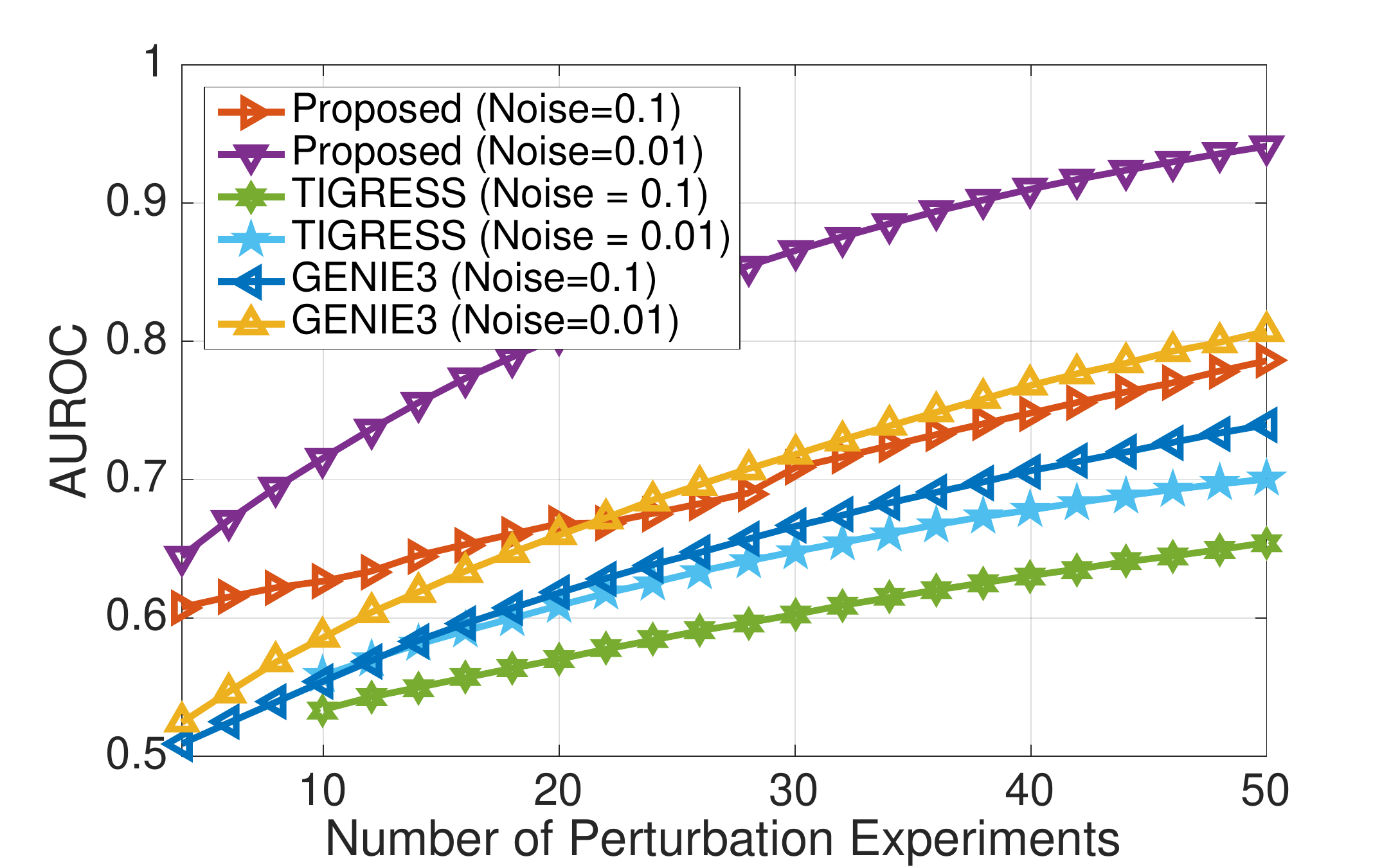}~
\includegraphics[width=0.45\linewidth]{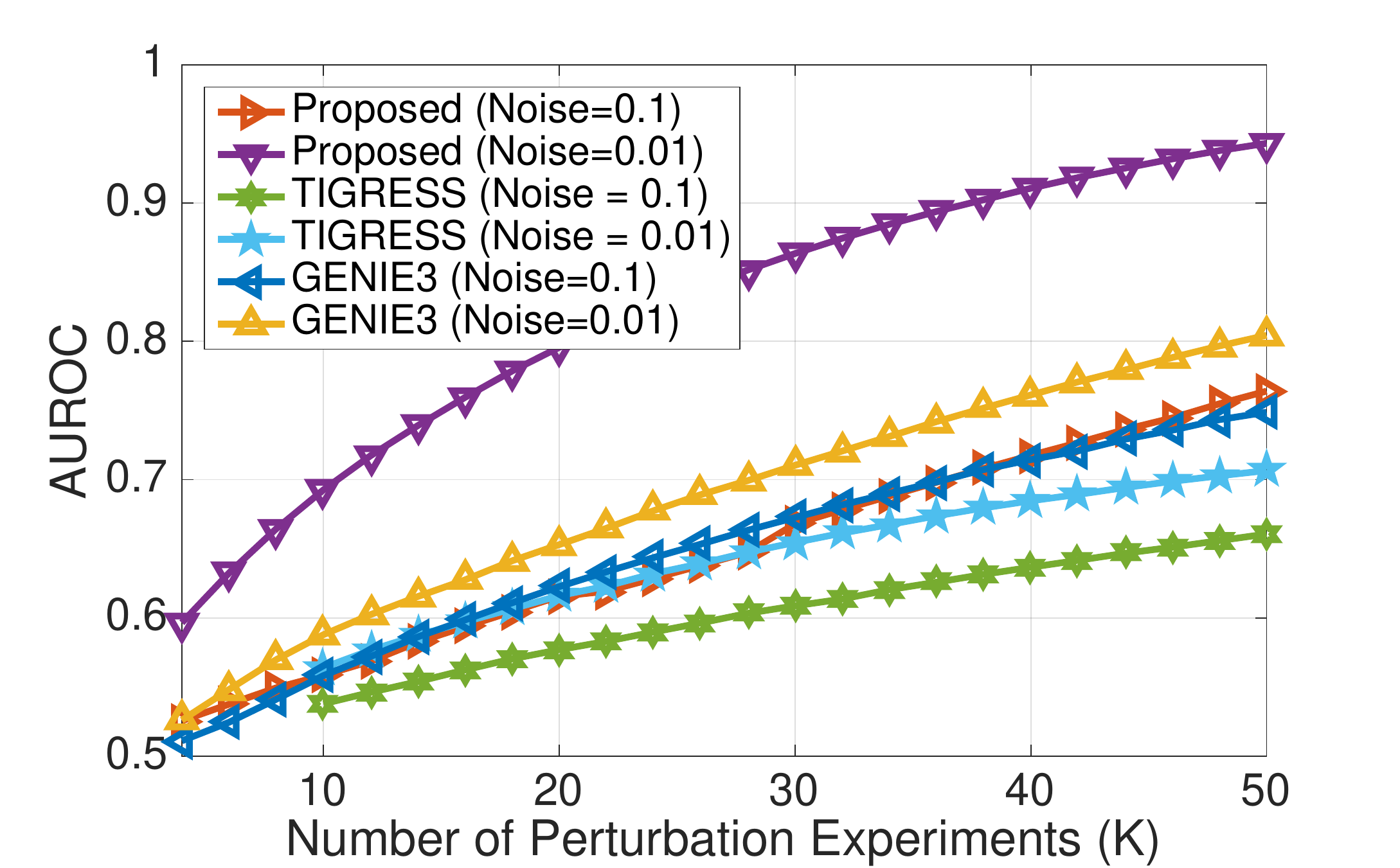}\vspace{.1cm}
\caption{\footnotesize Numerical Experiments on synthetic data of networks with $n=100$ nodes,
averaged over $100$ Monte-Carlo trials. 
Two random graph models are considered --- Erdos-Renyi graph with connectivity $p=0.1$ and 
Random regular graph with constant degree $d=10$. 
The model response function used is $h(x) = x^{0.5} / (1+x^{0.5})$ and the parameters are assumed
to be known.
The top figures consider the noiseless observation case. 
(Top-Left) Area under ROC (AUROC) against the number of perturbation experiments $K$. (Top-Right) 
Mean square error of the recovered $\hat{\bm A}$ using the proposed method.
Our proposed RIDS method (via solving \eqref{eq:sense}) 
outperforms GENIE3  and TIGRESS over all range of $K$ and achieves 
perfect recovery at $K \geq 32$ for the random regular graph model (as indicated by the MSE plot). 
Notice that the TIGRESS method has encountered numerical issues for the case with ER graphs.  
Perfect recovery was also observed for the ER model for $\sim 70\%$ of the instances at $K \geq 32$,
but however, as the ER graphs tend to have hubs with high degree,
its average MSE performance will be affected.   
The bottom figures consider the noisy observation case. 
(Bottom-Left) Random Erdos-Renyi graphs with connectivity $0.1$. (Bottom-Right) Random Regular graphs with constant degree $d=10$. Again, we observe that the proposed method has outperformed GENIE3 
and TIGRESS
for the considered range of $K$.\vspace{-0.1cm}
} \label{fig:syn} \label{fig:syn_noisy}
\end{figure}

\subsection{Synthetic Data} \label{sec:syn}\vspace{-.2cm}
\textbf{\emph{In silico} data}. We test the models when
the GRN $G = (V,E)$ with $n=100$.
The weight matrix
${\bm A}$ has entries that are uniformly distributed in $(0,1]$. 
We evaluate the steady-state 
gene expression levels
subject to gene deletion using the $4$th order Runge-Katta method. 
We include the GENIE3 method \cite{Huynh-Thu:2010aa} and 
TIGRESS method \cite{Haury:2012aa} implemented by their respective authors for benchmarks. 
The latter two were the best performing methods in the \texttt{DREAM5} challenge. 
All algorithms tested are implemented on MATLAB 2016a.
For the zeros index set ${\cal S}$ in \eqref{eq:mask}, we set $\delta = 0.02$. 

\noindent \textbf{Analysis}. 
Figure~\ref{fig:syn} (Top) compares the mean square error of the recovered $\hat{\bm A}$ 
and the area under an ROC curve (AUROC) for the recovered network $G$ versus the number of 
perturbation experiments $K$. We assume noiseless measurements in this case and 
solve \eqref{eq:sense} to recover the network.
We observe that the proposed method
achieves an AUROC of $\geq 0.9$ with $K \geq 14$ perturbation experiments,  
significantly outperforming the GENIE3 and TIGRESS methods under similar conditions. 
Moreover, we see that the proposed RIDS method has a better performance
when the underlying graph is a regular graph. In particular, with $K=32$ perturbation 
experiments
we have perfect recovery of both the inferred links and interaction strengths, $A_{ij}$, 
for the regular graph model. 
The above observation coincides with 
Theorem~\ref{thm:id} which predicts with $2.91 \times \max_{i \in [n]} \| {\bm a}_i \|_0$ 
perturbation experiments, one can perfectly recover the GRN with our proposed approach
when $n \rightarrow \infty$. 
Nevertheless, having $K \approx 14$ experiments
($\sim 15\%$ of the total number of genes)
is sufficient to yield a good GRN reconstruction performance. 

Figure~\ref{fig:syn_noisy} (Bottom) considers the \emph{noisy} measurement scenario. 
With reference to \eqref{eq:measure}, the elements 
of $\overline{\bm{\epsilon}}, \bm{\epsilon}[k]$ are independently extracted from 
normal distributions ${\cal N}(0,0.1)$ and ${\cal N}(0,0.01)$. 
We apply the robust formulation \eqref{eq:sense_r} with the regularizing parameters set to 
$\rho = 10^{-5}, \gamma = 0.5$, $\gamma = 0.05$ for the case with ${\cal N}(0,0.1)$ noise and ${\cal N}(0,0.01)$ noise, respectively, to recover the network, notice that in this scenario
${\bm b}$ is known and problem \eqref{eq:sense_r} is solved directly. 
Comparing the average AUROC performance shows that 
our method has  a consistently better
performance than GENIE3 and TIGRESS. 
However, as the noise power grows, the advantage of applying our method 
declines.
\vspace{-.3cm}

\subsection{Empirical Data} \label{sec:dream5_r}\vspace{-.2cm}
\textbf{\emph{In vivo} data}. To test our methodology against empirical data, we focus on the reconstruction performance of
{\it Escherichia coli} ({\it E. coli}) and {\it Saccharomyces cerevisiae} ({\it S. cerevisiae})  from 
gene perturbation experiments, using 
the highly curated datasets collected for the \texttt{DREAM5} Network Inference Challenge\footnote{Available at \verb|https://www.synapse.org/#\!Synapse:syn2787209/wiki/70351|.}. 
The \emph{E. coli} (\resp \emph{S. cerevisiae}) dataset describes 
$805$ (\resp $536$) vectors of expression level of $n=4511$ 
(\resp $n=5950$) anonymized genes under 
 different experimental conditions; 
the dataset also lists a subset of $TF = 334$ (\resp $TF=333$) genes that are
recognized as known \emph{transcription factors} (TFs), some of whom are decoys, namely
wrongly labeled as such.

Our method relies on the expression levels in the steady state only. 
Therefore, we use only $326$ out of $805$ (\resp $238$ out of $536$) vectors of expression levels 
in the \emph{E. coli} (\resp \emph{S. cerevisiae}) dataset, \ie about $\sim 40\%$ of the data.  
The remaining vectors correspond to transient states, which 
our method is not designed to treat. 
Upon grouping and denoising the dataset using 
the pre-processing method in Appendix~\ref{sec:dream5},
we are left with $K=56$ (\resp $K=7$) vectors of expression levels 
for \emph{E. coli} (\resp \emph{S. cerevisiae}),
each with a distinct perturbation conditions. 
From these vectors, we must reconstruct $\bm A$, capturing the GRN
between the transcription factors and the genes, a total of 
$334 \times 4511$ (\resp $333 \times 5950$) potential links for \emph{E. coli} (\resp \emph{S. cerevisiae}). 

We use the following model response function for empirical data: 
\beq \label{eq:template}
h(x; {\bm b}) = 0.75 \cdot x^{b_2} / (1 + b_1 x^{b_2} ),~{\cal B} = \{ {\bm b} \in \RR^2~:~b_1, b_2 \geq 0\} \eqs.
\eeq
The above encompasses several kinetic interaction models. 
When $b_1 \rightarrow 0$, the interaction model tends to be that of a chemical activation process;
otherwise the interaction model has a saturating effect close to a switch-like process; 
$b_2$ controls the saturation rate of the gene interaction.
We apply the RIDS method to learn simultaneously the GRN and the above
model parameters. 
Further details of the experimental procedure can be found in Appendix~\ref{sec:dream5}. 

{\renewcommand{\arraystretch}{1.05}
\newcolumntype{C}[1]{>{\centering\let\newline\\\arraybackslash\hspace{0pt}}m{#1}}

\begin{table}[!t]
\begin{center}
{\footnotesize \sf \begin{tabular}{| c|| C{1.8cm} | C{1.7cm} | C{1cm} || C{1.7cm} | C{1.7cm} | C{1cm} |} 
\hline
& \multicolumn{3}{c||}{\emph{E. coli}} & \multicolumn{3}{c|}{\emph{S. cerevisiae}} \\
\hline
Methods & AUROC & AUPR & Score & AUROC & AUPR & Score \\
\hline
\hline
TIGRESS \cite{Haury:2012aa} & 0.595 & 0.069 & 4.41 & 0.517 & 0.02 & 1.082 \\
\hline
GENIE3 \cite{Huynh-Thu:2010aa} & 0.617 & \bfseries 0.093 & 14.79 & 0.518 & 0.021 & 1.387 \\
\hline
RankSum & 0.65 & 0.09 & 24.90 & \bfseries 0.528 & 0.022 & 6.236 \\
\hline
bLARS \cite{singh16} & N/A & N/A & 5.841 & N/A & N/A & \bfseries 7.479 \\
\hline
\emph{RIDS} & 0.6808 & 0.0504 & 32.39 & 0.525  & \bfseries 0.022 & 4.694 \\
\emph{(top 100k)}	& $1.69\times 10^{-64}$ & $9.9 \times 10^{-2}$ & & $3.84 \times 10^{-8}$ & $2.3 \times 10^{-2}$ & \\
\hline
\emph{RIDS (opt.~${\bm b},$} & \bfseries 0.6823 & 0.0508 & \bfseries 33.29 & 0.525 & 0.021 & 4.161 \\
		\emph{top 100k)}   &  $ 3.13\times 10^{-66}$ & $8.5 \times 10^{-2}$ & & $6.47 \times 10^{-8}$ & $2.9 \times 10^{-2}$ & \\
\hline
\emph{RIDS (no TF,} & 0.6745 & 0.0540 & 29.12 & 0.524 & 0.0221 & 4.298 \\
		\emph{top 100k)}   & $2.78\times 10^{-57}$ & $2.0 \times 10^{-2}$ & & $3.70 \times 10^{-7}$ & $4.8 \times 10^{-2}$ &  \\
\hline
\hline
iRafNet \cite{Petralia:2015aa} & 0.641 & \bfseries 0.112 & 29.26 & N/A & N/A & N/A \\
\hline
GENIMS \cite{Wu:2016aa} & 0.705 & 0.052 & 48.33 & 0.533 & 0.02 & 8.454 \\
\hline
\emph{RIDS} & \bfseries 0.7573 &  0.0574 & \bfseries 93.28 & \bfseries 0.5734 & \bfseries 0.0252 & \bfseries 62.64 \\
\emph{(top 500k)}           & $1.04\times 10^{-184}$ & $2.7 \times 10^{-3}$ & & $1.5 \times 10^{-119}$ & $2.38 \times 10^{-7}$ & \\
\hline 
\end{tabular}} \vspace{-.1cm}
\end{center}
{\sf \footnotesize $^\star$All values/scores are calculated with the top 100k predictions.
Exceptions are the iRafNet, GENIMS and RIDS (top 500k) in the last three rows, that are based on the top 200k, all, top 500k predictions, respectively.}
\caption{\footnotesize GRN recovery result on the two  \emph{in vivo}
dataset. Scores for RankSum, GENIE3 and TIGRESS are taken directly from the
{DREAM5}'s leaderboard. 
Notice that RankSum is the community integrated prediction.
For the RIDS method, the lower scores are the $p$-value for the AUROC/AUPR metrics. 
The RIDS method in the $5$th row uses the median optimized parameters learnt for the two networks, \ie  $b_1 = 0.047, b_2 = 0.5893$ for \emph{E. coli} and $b_1 = 0.5571, b_2 = 0.3749$ for \emph{S. cerevisiae} and we solve \eqref{eq:sense_r} with the fixed parameters for all genes; the $6$th row corresponds 
to tackling \eqref{eq:sense_r} \emph{without} using the list of transcription factors.
Our method obtains the best prediction score for \emph{E. coli},
and the score for \emph{S. cerevisiae} is comparable to state-of-the-art.} \vspace{-.4cm}\label{tab:result}
\end{table}
}

\noindent \textbf{Analysis}. Table~\ref{tab:result} shows the GRN recovery result for the
truncated top 
\emph{100,000}
and \emph{500,000} 
predicted links in the GRN, compared to the Gold Standard. 
For each method, we evaluate the AUROC, AUPR and the prediction score,
defined similarly as in  \cite{Marbach:2012aa}
by ${\sf Score} \eqdef ( \log_{10}( p_{AUROC} ) + \log_{10} ( p_{AUPR} )) / 2$,
where $p_{AUROC}$ and $p_{AUPR}$ are the $p$-values for AUROC/AUPR. 
The numerical experiment demonstrates that our robust
GRN recovery approach 
is able to infer GRN from \emph{few steady-state data}, 
while achieving superior performance to the state-of-the-art. 
In particular, for the \emph{E. coli} network, the RIDS method is the top performer among the 
compared methods, beating even the community integrated prediction (RankSum)
while using $\sim 60\%$ less experimental data. 
Our method gives good performance even the list of TFs are not provided 
when tackling \eqref{eq:sense_r}.  

\begin{figure}[!t]
\centering
\includegraphics[width=0.4\linewidth]{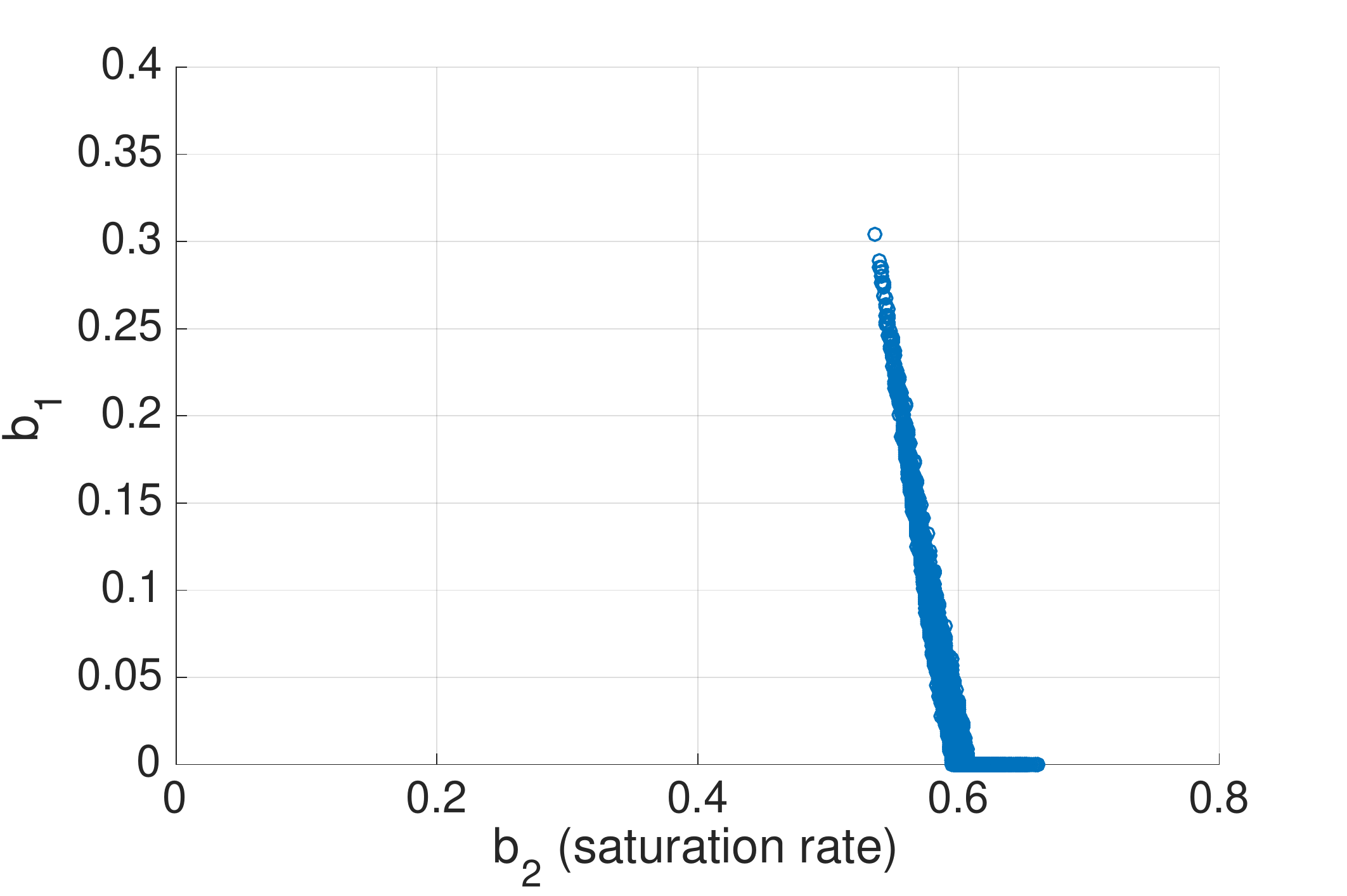}
\includegraphics[width=0.4\linewidth]{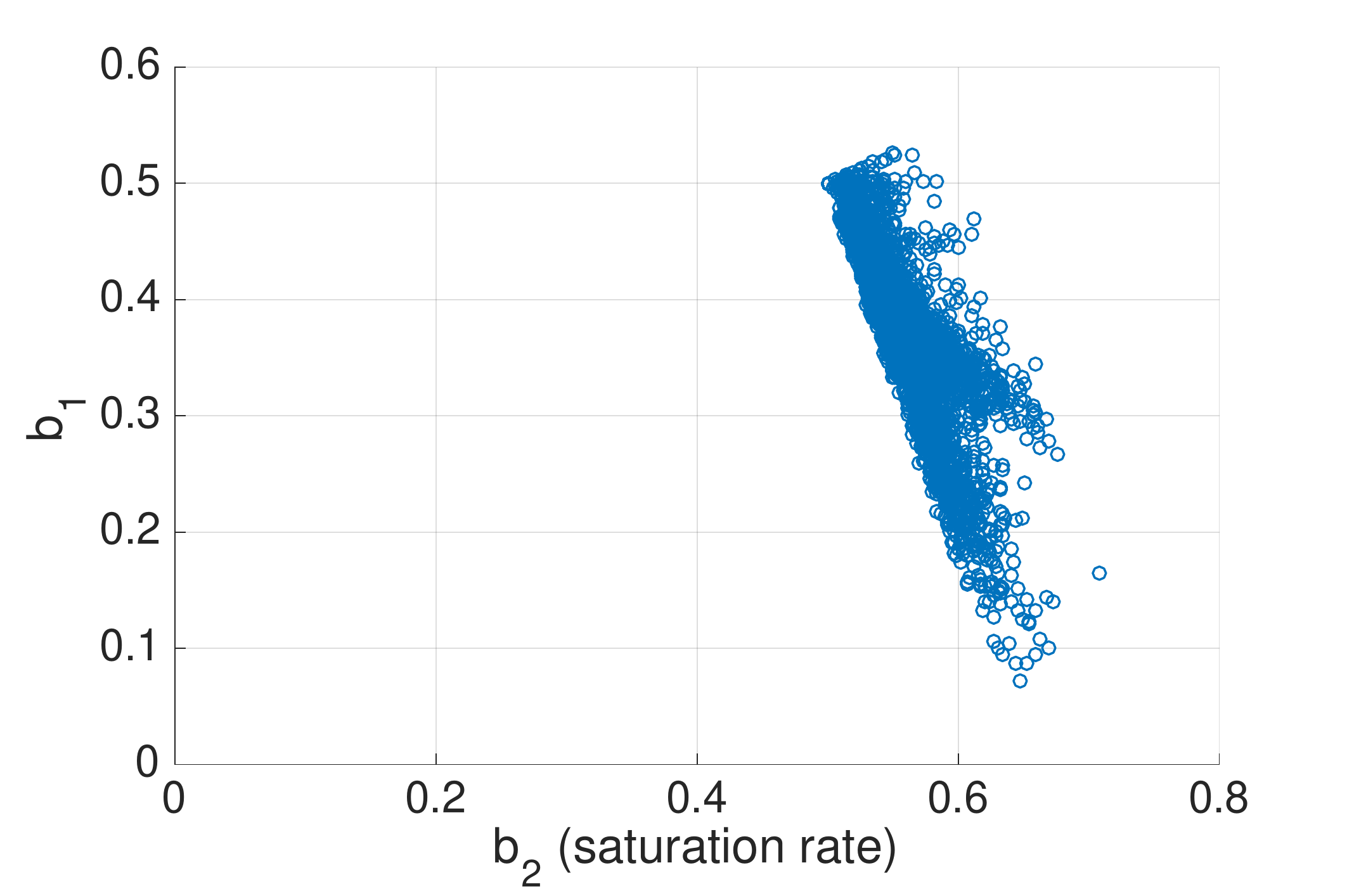}\vspace{-.1cm}
\includegraphics[width=0.4\linewidth]{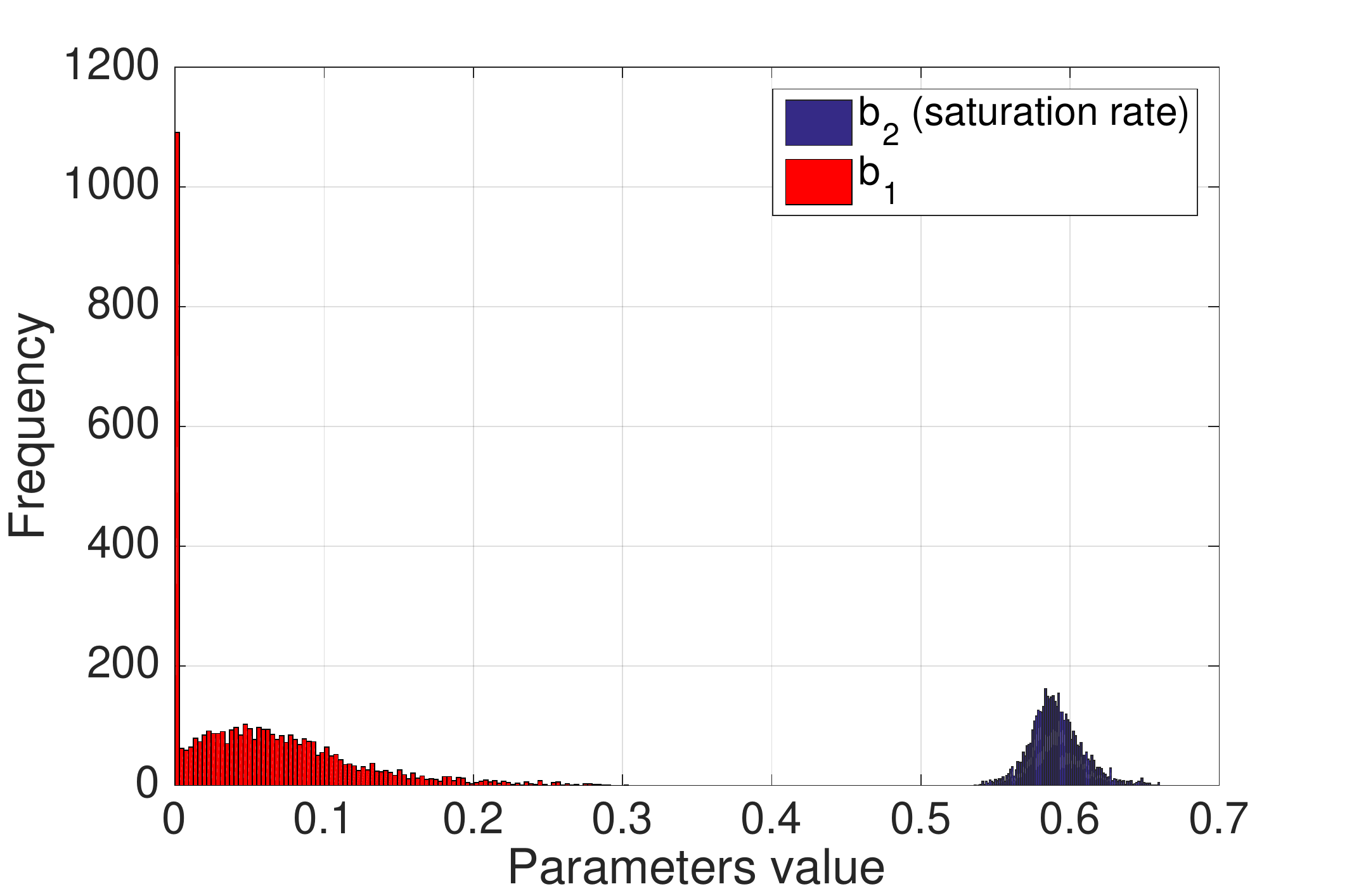}
\includegraphics[width=0.4\linewidth]{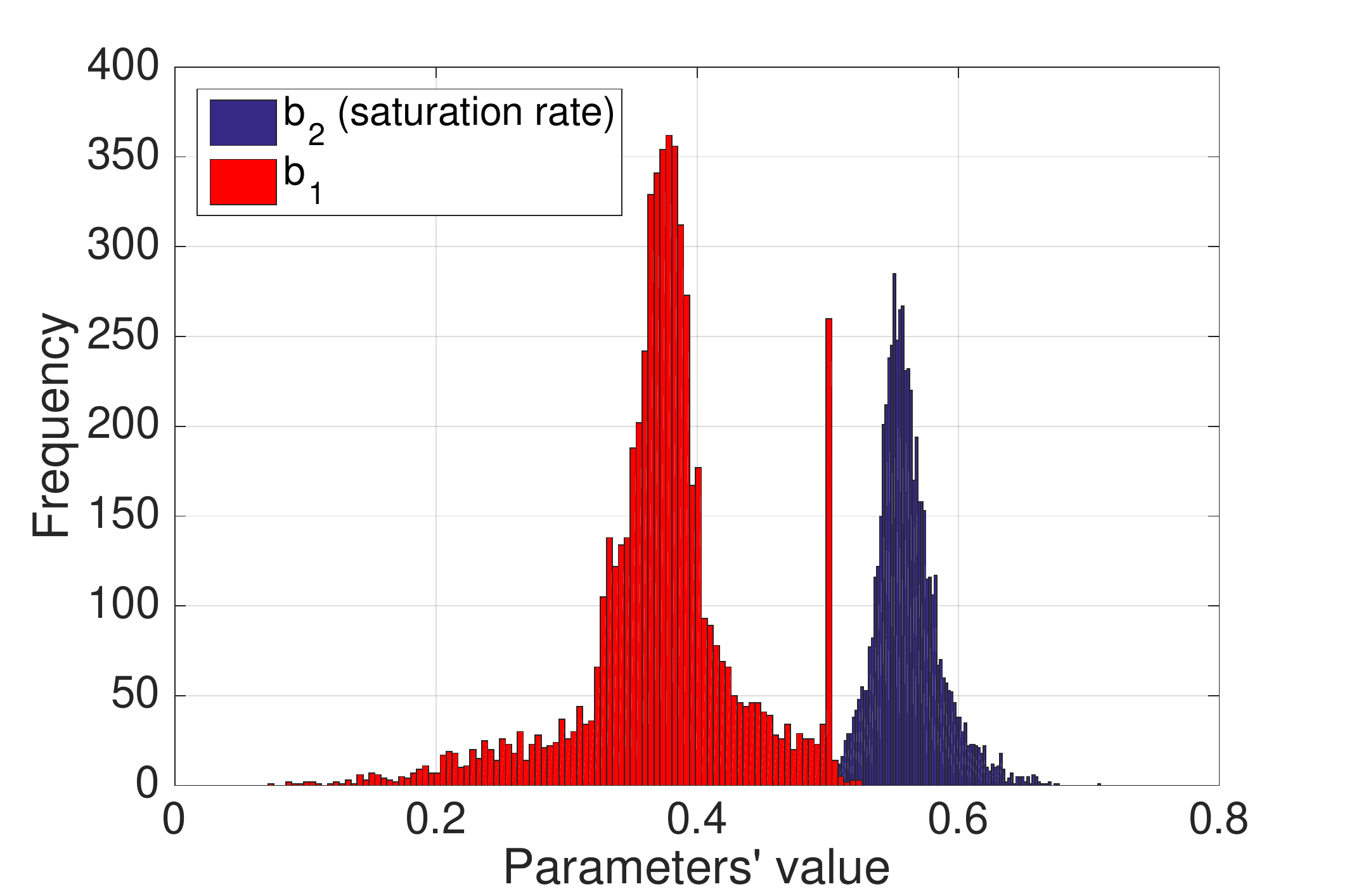}\vspace{-.3cm}
\caption{\footnotesize Model parameters learnt for different networks. (Left) \emph{E. coli} network. (Right) \emph{S. cerevisiae} network. The top figures show the scatter plot of the parameters. 
The bottom figures show the histograms 
of parameters learnt, where the {\color{blue} blue} patch is the 
saturation rate parameter $b_2$ and the {\color{red} red} patch is parameter $b_1$ which 
controls if the interaction model is more of the chemical activation type or the switching type, cf.~\eqref{eq:template}. 
Meanwhile, the \emph{S. cerevisiae} shows 
a higher variance for the $b_1,b_2$ values learnt for each gene, this is due to the fact
that the in vivo data available for this network is more scarce than the one for \emph{E. coli}. 
\vspace{-.4cm}} \label{fig:param}
\end{figure}

The model parameters learnt using RIDS are plotted
in Figure~\ref{fig:param}. We observe that the parameters learnt are clustered around 
$b_1 \sim 0.05$, $b_2 \sim 0.6$ for the \emph{E. coli} network and $b_1 \sim 0.4$, $b_2 \sim 0.55$ for the \emph{S. cerevisiae} network. Suggesting that the interaction model for the former is close to a chemical activation process, while the latter is close to being
a combination. 
The parameters learnt are also similar across different genes, which seems to indiciate that  the 
parameter space may be reduced by using a single set of parameters for all genes. 
This is corroborated by the AUROC/AUPR performance obtained
when the RIDS method 
is re-applied with the same model parameter for all genes in Table~\ref{tab:result}. 
Additional results on the empirical data can be found in Appendix~\ref{app:add},
where we further demonstrate that RIDS can infer directionality from empirical GRNs.

\noindent \textbf{Computational Complexity}. 
The problem \eqref{eq:sense} and the AO procedure in \eqref{eq:ao} can be
solved by applying generic optimization
packages such as \texttt{cvx}. 
It takes about 25 seconds to recover the synthetic networks in Section~\ref{sec:syn} with
$n=100$ genes on a quad-core labtop computer 
and  about $1100$ seconds for the $n=4511$ genes \emph{E.~Coli.} 
network in Section~\ref{sec:dream5_r} on a 60-cores server.
Further speed up will be possible if we apply specialized optimization tools. \vspace{.0cm}

\noindent \textbf{Conclusions}.
This work proposes the RIDS method for GRN inference
using perturbation experiments. 
The RIDS method is developed through modelling the gene expression
data as the outcome of perturbing a nonlinear dynamic system. 
To improve robustness, the method first applies a 
subspace projection method for denoising the gene expression data,
then a sparse and robust estimator is applied to recover the network.
The model parameters of the dynamic system will also be inferred simultaneously. 
Our theoretical analysis, conducted under the assumption
that the dynamic's parameters are known, shows that it is possible to recover the GRN even
when there are only a few sets of perturbation experiment data available. This 
is in contrary to the common belief that it requires a large number of 
experiments to apply similar methods.
Moreover, our experiments on empirical data shows that 
the RIDS method compares favorably to the state-of-the-art methods
while requiring $\sim 60\%$ less data. 
The RIDS method paves the way to study the dynamics
of gene interactions by having the ability to infer the model parameters,
which is important as studied by \cite{Ronen:2002aa}. 
For example, our preliminary result suggests that the genes in the 
\emph{E. coli} and \emph{S. cerevisiae} networks tend to have a different 
interaction model with its neighboring genes.

\newpage
{\small 
\bibliography{gene}
}

\newpage

\appendix

\section{Proof of Proposition~1}\label{app:pert}
Define
\beq
\olx - {\bm x}[k] = ([\olx]_k-z_k) {\bf e}_k + \bm{\epsilon},
\eeq
such that the $k$th component of $\bm{\epsilon}$ is zero. Our goal is to find $\bm{\epsilon}$. We have:
\beq \label{eq:subtract}
\olx - {\bm x}[k]  = {\bm A} ( {\bm h}( \olx ) - {\bm h}( {\bm x}[k] ) ) + {\bf e}_k {\bf e}_k^\top {\bm A} {\bm h}({\bm x}[k]) + z_k {\bf e}_k.
\eeq
Notice that ${\bf e}_k$ is in the null space of $({\bm I} - {\bf e}_k {\bf e}_k^\top )$. Left-multiplying $({\bm I} - {\bf e}_k {\bf e}_k^\top )$ to the both sides of the equation yields:
\beq
\begin{split}
\bm{\epsilon} & = ({\bm I} - {\bf e}_k {\bf e}_k^\top ) {\bm A} ( {\bm h} ( \olx ) - {\bm h} ( {\bm x}[k] ) ) \\
&  \approx ({\bm I} - {\bf e}_k {\bf e}_k^\top ) {\bm A} \grd {\bm h}( {\bm x}[k]) ( ([\olx]_k-z_k) {\bf e}_k + \bm{\epsilon} ) 
\end{split}
\eeq
where we have taken the Taylor's expansion for ${\bm h} (\olx )$ centered at ${\bm x}[k]$ and the 
approximation is accurate when the perturbation $\olx - {\bm x}[k]$ is small. 
This gives
\beq \label{eq:obs}
\bm{\epsilon} \approx ({\bm I} - {\bf e}_k {\bf e}_k^\top ) {\bm A} \grd {\bm h}( {\bm x}[k]) \bm{\epsilon} + ([\olx]_k-z_k) ({\bm I} - {\bf e}_k {\bf e}_k^\top ) {\bm A} \grd {\bm h}( {\bm x}[k]) {\bf e}_k
\eeq
Moreover, we notice that $\grd {\bm h}( {\bm x}[k])$ is a diagonal matrix with $h'(z_k)$ on its $k$th entry, therefore the latter term can be simplified as
\beq \begin{split}
 ([\olx]_k-z_k) ({\bm I} - {\bf e}_k {\bf e}_k^\top ) {\bm A} \grd {\bm h}( {\bm x}') {\bf e}_k & =  ([\olx]_k-z_k)  h'(z_k) ({\bm I} - {\bf e}_k {\bf e}_k^\top ) {\bm a}_{col,k} \\
 & =  ([\olx]_k-z_k)  h'(z_k) {\bm a}_{col,k} \eqs,
 \end{split}
\eeq
where the last equality is due to the fact that the $k$th element of ${\bm a}_{col,k}$ is zero. 

Finally, we observe that
\beq 
\begin{split}
\bm{\epsilon} & = \big( {\bm I} - ({\bm I} - {\bf e}_k {\bf e}_k^\top ) {\bm A} \grd {\bm h}( {\bm x}[k])  \big)^{-1} \cdot ([\olx]_k-z_k) h'(z_k) {\bm a}_{col,k} \\
& = \big({\bm I} + ({\bm I} - {\bf e}_k {\bf e}_k^\top ) {\bm A} \grd {\bm h}( {\bm x}')  \\
& \hspace{2cm} + (({\bm I} - {\bf e}_k {\bf e}_k^\top ) {\bm A} \grd {\bm h}( {\bm x}'))^2 + \cdots \big) \cdot ([\olx]_k-z_k) h'(z_k) {\bm a}_{col,k} \\
& \approx ([\olx]_k-z_k) h'(z_k) {\bm a}_{col,k} \eqs,
\end{split}
\eeq
where the second equality is due to the Taylor's expansion. 
Notice that the series expansion holds when $\lambda_{max} ( ({\bm I} - {\bf e}_k {\bf e}_k^\top ) {\bm A} \grd {\bm h}( {\bm x}') ) < 1$. \hfill 

\section{Proof of Theorem~1} \label{app:id}
\textbf{Proof Outline}. Using the assumptions stated in the Theorem, we first reduce the linear
system into a simple form that involves an underdetermined system with 
a \emph{sparse} sensing matrix. The sensing matrix is then found to have the 
same structure/support as the bipartite graph formed by the edges 
\emph{from} the perturbed nodes \emph{to} all other nodes. Finally, the 
desired perfect recovery condition is  
given as a consequence of the expander graph property of this bipartite graph.

Let us give a formal definition of expander graph:
\begin{Def} \label{def:exp}
An $(\alpha,\delta)$-unbalanced expander graph, $G(A,B)$, is an $A,B$-bipartite graph (bigraph) with $|A| = n, |B| = m$ with left degree bounded in $[d_l, d_u]$, i.e., $d(v_i) \in [d_l, d_u]$ for all $v_i \in A$, such that for any $S \subseteq A$ with $|S| \leq \alpha n$, we have $\delta |E(S,B)| \leq |N(S)|$, where $E(S,B)$ is the set of edges from  $S$ to $B$ and $N(S) = \{ v_j \in B : \exists~v_i \in S~s.t.~(v_j, v_i) \in E \}$ is the neighbor set of $S$ in $B$.
\end{Def}

We are ready to begin our proof. 
Let ${\bm a}_i$ be the $i$th row of ${\bm A}$ and define a set of vectors $\{ {\bm y}_i \}_{i=1}^n$
such that $[{\bm y}_i]_k \eqdef \overline{x}_i - x_i[k]$. 
We observe that ${\bm y}_i$ is a collection of the data points that depend on ${\bm a}_i$.
For simplicity, let us focus on the case when $i \notin [K]$,
\beq
[{\bm y}_i]_k = {\bm a}_i^\top ( {\bm h} ( \olx ) - {\bm h} ( {\bm x}[k] ) ) 
\approx ( \overline{x}_k - z_k ) \cdot {\bm a}_i^\top  \grd {\bm h}(\olx) \big( h'(z_k) {\bm a}_{col,k} + {\bf e}_k \big) \eqs,
\eeq
where we have applied Proposition~\ref{prop:pert} and the first order Taylor approximation on 
${\bm h} ( \olx ) - {\bm h} ( {\bm x}[k] )$ to yield the result above. 
Notice that we have dropped the dependence on ${\bm b}$ as the ODE parameter
is assumed to be known in this setting. 
After some manipulations and applying the assumptions stated in the theorem, we can 
express the equation above as
\beq \label{eq:mm_ei}
{\bm y}_i = \left( \begin{array}{cc} 
\bm{\Lambda} &  {\bm 0}_{K \times (n-K)} 
\end{array} 
\right) {\bm a}_i + 
\underbrace{ \left( \begin{array}{c} 
( \overline{x}_1 - z_1) h'(z_1) \cdot \grd {\bm h}(\olx) {\bm a}_{col,1}^\top \\
\vdots \\
( \overline{x}_K - z_K ) h'(z_K)  \cdot \grd {\bm h}(\olx) {\bm a}_{col,K}^\top \\
\end{array}\right)}_{\eqdef \tilde{\bm E}_i} {\bm a}_i \eqs,
\eeq
where $\bm{\Lambda}$ is an $K \times K$ diagonal matrix with the $k$th element being 
$[\bm{\Lambda}]_{kk} = \overline{x}_k - z_k$. 
The challenge is that as the non-zero elements of $\bm{\Lambda}$ has a larger magnitude 
than the matrix $\tilde{\bm E}_i$ in the latter matrix-vector product, the overall
sensing matrix $( \bm{\Lambda}~ {\bf 0} ) + \tilde{\bm E}_i$ is dominated 
by the diagonal matrix component. That is, 
$( \bm{\Lambda}~ {\bf 0} ) + \tilde{\bm E}_i \approx ( \bm{\Lambda}~ {\bf 0} )$. 
However, this implies that the elements of ${\bm a}_i$ over the coordinates $[n] \setminus [K]$
cannot be recovered from ${\bm y}_i$ as the rows of the sensing matrix supported only on $[K]$. 
It is thus necessary to exploit extra information to recover ${\bm a}_i$. 

In light of this, we note that $[{\bm a}_i]_j$ is zero for all $j$ in ${\cal S}_i$. 
The first matrix-vector product is thus an $K-|{\cal S}_i|$-sparse vector which is supported on the 
set ${\cal S}_i^c = [K] - {\cal S}_i$.
As we are interested in studying a sufficient condition for perfect recovery,
we see that \eqref{eq:mm_ei} implies the following linear system with
 the rows corresponding to ${\cal S}_i^c$ removed, 
\beq \label{eq:mm_support}
\big[ {\bm y}_i \big]_{ {\cal S}_i } = 
\big[ \tilde{\bm E}_i \big]_{ {\cal S}_i, : } {\bm a}_i \eqs. 
\eeq
Compared to the original model \eqref{eq:mm_ei}, 
we can suppress the dominating diagonal component in 
$\tilde{\bm E}_i$ using the support knowledge on ${\bm a}_i$.

The remaining task is to verify if the reduced sensing matrix 
$\big[ \tilde{\bm E}_i \big]_{ {\cal S}_i, : }$ is a good sensing matrix. 
Notice that \emph{dense} and random 
matrices are known to exhibit good properties in which one only requires  
$|{\cal S}_i| \geq 2 \| {\bm a}_i \|_0 \cdot \log n$ to achieve perfect recovery. 
On the other hand, $\big[ \tilde{\bm E}_i \big]_{ {\cal S}_i, : }$ is a \emph{sparse} matrix whose
support depends on the out-neighbors of the nodes in ${\cal S}_i$. In particular, we have ${\rm supp} ( \big[ \tilde{\bm E}_i \big]_{ {\cal S}_i, : } ) = {\rm supp} ( {\bm A}_{:, {\cal S}_i}^\top)$. 
An interesting observation is that the support of the sensing matrix depends on the support of 
${\bm A}$ itself or the network that we wish to recover. 

As it turns out, the perfect recovery condition for ${\bm a}_i$ boils down to
studying conditions on the \emph{support} of 
$ \big[ \tilde{\bm E}_i \big]_{ {\cal S}_i, : }$. 
In particular, let $G_{bi}(A,B)$ with $|A| = n$ and $|B| = |{\cal S}_i| \leq K$ be the bi-partite graph representation of 
$[ {\bm E}_i ]_{{\cal S}_i, :}$ such that the adjacency 
matrix of $G_{bi}$, i.e., ${\bm A}_{bi} \in \RR^{|{\cal S}_i| \times n}$, has the same support as 
$[ {\bm E}_i ]_{{\cal S}_i, :}$. 

\begin{Theorem} \label{thm:exp}
If $G_{bi}$ is an $(\alpha, \delta)$-unbalanced expander graph with left degree bounded in $[d_l, d_u]$ such that $2 (d_l / d_u) \cdot \delta > \sqrt{5} - 1$ and $k \leq \frac{\alpha}{1+\rho \delta} n$, and $\tilde{\bm E}_i$ or $-\tilde{\bm E}_i$ is non-negative, then the set 
\beq
{\cal C} = \{ \hat{\bm a}_i~:~ \big[ \tilde{\bm E}_i \big]_{ {\cal S}_i, : } ( \hat{\bm a}_i - {\bm a}_i ) = {\bm 0} \}
\eeq
is a singleton for all $k$-sparse vector ${\bm a}_i$. 
\end{Theorem}

The proof of the theorem above can be found in subsection B.2. 
Consequently, we observe that all the conditions in Theorem~\ref{thm:exp} hold, 
therefore solving \eqref{eq:sense} yields $\hat{\bm A} = {\bm A}$, \ie 
we achieve perfect recovery of the network. 

\subsection{A special case satisfying the conditions in Theorem~\ref{thm:id}} \label{app:id_insight}

Let us first
borrow the following proposition from \cite{tsipn_submit}:
\begin{Prop} \label{prop:exp} \cite[Proposition 5]{tsipn_submit}
Let $G(A,B)$ be a random bipartite with $|A| = n$, $|B| = \beta \cdot n$ for some $\beta < 1$ and 
the left degree bounded in $[d_l,d_u]$ (\ie nodes in $A$ has bounded degree). 
Fix $\alpha \in (0,1)$ and the following holds: 
\beq \label{eq:highprob}
d_l \geq 3,~~\beta > \alpha,~~d_l > ( H( \alpha ) + \beta H( \alpha / \beta ) ) / (\alpha \log ( \beta / \alpha ) ) \eqs,
\eeq
where $H(\alpha) = \alpha \log \alpha + (1-\alpha) \log(1 - \alpha)$ is the binary entropy function. 
Then, with high probability as $n \rightarrow \infty$, the graph
$G(A,B)$ is an $(\alpha, 1-1/d_l)$-expander.
\end{Prop}
Following our discussions in Section~\ref{sec:theory}, we 
suppose that the perturbed genes $[K]$ are chosen
such that for each $j \in V$, the $j$th gene has $d$ in-neighbors in $[K]$. 
Notice
that this requires selecting perturbation nodes strategically, e.g., using advices from the experts. 

Furthermore, the bipartite graph of the $K \times n$ matrix $([{\bm A}]_{:, [K]})^\top$
may follow a regular random graph model with constant degree $d$. 
Then, using similar arguments as in \cite[Proposition 4]{tsipn_submit},
it can be shown that 
the support of the sub-matrix $([{\bm A}]_{:, {\cal S}_i})^\top$ corresponds to a random bipartite 
graph with bounded degree in $[d-1,d]$ with high probability.
The intuition is that the matrix $([{\bm A}]_{:, {\cal S}_i})^\top$ is
formed by deleting a small number of rows from 
the $K \times n$ matrix $([{\bm A}]_{:, [K]})^\top$. 

Now fix $\alpha \in (0,1)$. 
Applying Proposition~\ref{prop:exp} shows that
there exists $\beta = c(d) \cdot \alpha$ with $c(d) > 1$ such that this bipartite 
graph is an $(\alpha, 1-1/(d-1))$-expander 
with high probability and $\beta$ satisfying the conditions in \eqref{eq:highprob}. 
Taking $|{\cal S}_i| = \beta n > c(d) \cdot \alpha  n$, we can substitute the numbers into 
the conditions \eqref{eq:id_cond}
required by Theorem~\ref{thm:id} to obtain an upper bound on $\| {\bm a}_i \|_0$
where we can recover
${\bm A}$ perfectly. This yields the desirable result in \eqref{eq:highlv}.

\subsection{Proof of Theorem~\ref{thm:exp}}
To prove Theorem~\ref{thm:exp}, the following Lemma would be instrumental. Denote
${\rm Null}( {\bm E})$ as the null space of a matrix ${\bm E}$,
$S_-({\bm w}) = \{ i \in [n] : w_i < 0\}$ and 
$S_+({\bm w}) = \{ i \in [n] : w_i > 0\}$ be the negative and positive support of
${\bm w}$. 
\begin{Lemma} \label{lem:thm2}
If (i) $\overline{\bm x}$ is $k$-sparse, (ii) ${\bm E} \in \mathbb{R}^{m \times n}$ satisfies that ${\bm 0} \neq {\bm w} \in {\rm Null}({\bm E})$, $|S_-({\bm w})| \geq k+1$, then the set
${\cal C} = \{ {\bm x} : {\bm E} ({\bm x} - \overline{\bm x} ) = {\bm 0},~{\bm x} \geq {\bm 0} \}$ is a singleton.
\end{Lemma}

\noindent \textbf{Proof}. Suppose $|{\cal C}| > 1$ such that  there exists $\tilde{\bm x} \in {\cal C}$, $\tilde{\bm x} \neq \overline{\bm x}$. It is straightforward to show that $\tilde{\bm x} = \overline{\bm x} + {\bm w}$, where ${\bm w} \in {\rm Null}({\bm E})$. The assumption implies that $|S_-({\bm w})| \geq k+1$ and $\tilde{\bm x} = \overline{\bm x} + {\bm w} \ngeq {\bm 0}$ as $\overline{\bm x}$ is $k$-sparse. This contradicts $\tilde{\bm x} \in {\cal C}$. \hfill {\bf Q.E.D.}

The next step is to apply a generalization of \cite[Theorem~4]{Wang2011}:
\begin{Lemma} \label{lem:single}
Let $n > m$ and ${\bm E} \in \mathbb{R}^{m \times n}$ be a non-negative matrix that has the same support as the adjacency matrix of an $(\alpha,\delta)$-unbalanced bipartite expander graph with left degrees bounded 
in $[d_l, d_u]$ and $\rho = d_l / d_u$. If $\rho \delta > (\sqrt{5} - 1)/ 2$, then for all $k$-sparse vector $\overline{\bm x}$. The set
\begin{equation}
{\cal C} = \{ {\bm x} : {\bm E} ({\bm x} - \overline{\bm x} ) = {\bm 0},~{\bm x} \geq {\bm 0} \},
\end{equation}
is a singleton if $k \leq \frac{ \alpha } {1 + \rho \delta } n$.
\end{Lemma}

\noindent \textbf{Proof}. 
Using Lemma~\ref{lem:thm2}, it suffices to prove that any ${\bm w} \in {\rm Null}({\bm E})$ with $S_- ({\bm w}) \subseteq A$, we have $|S_- ({\bm w})| \geq k+1$.
We shall proceed by contradiction. Suppose that there exists ${\bm w} \in {\rm Null}({\bm E})$ such that $|S_- ({\bm w})| \leq k$. Since $|S_- ({\bm w})| \leq k \leq \alpha n$, the expander property implies:
\begin{equation} \label{eq:same_chain}
\delta d_l \cdot |S_-({\bm w})| \leq \delta |E(S_-({\bm w}),B)| \leq |N(S_- ({\bm w}))| \eqs,
\end{equation}
where $N(S)$ denotes the set of neighbors of $S$ and $E(A,B)$ is the set
of edges between $A$ and $B$. 
The right hand side can be further upper bounded as:
\begin{equation} 
|N(S_- ({\bm w}))| \leq |E(S_-({\bm w}), B)| \leq d_u \cdot |S_- ({\bm w})|
\end{equation}
Moreover, we know that $N(S_-({\bm w})) = N(S_+({\bm w})) = N(S_-({\bm w}) \cup S_+({\bm w}))$. This is because ${\bm E} {\bm w} = {\bm 0}$ and ${\bm A}$ is non-negative, thus the neighborhood sets must coincide to enforce nullity, otherwise this will result in ${\bm E} {\bm w} \neq {\bf 0}$. Note that in \cite{Wang2011}, the proof is achieved by assuming that ${\bm E}$ is the binary adjacency matrix. We extend the same argument to the case when ${\bm E}$ is non-negative. Using the above discussions and the inequality \eqref{eq:same_chain} applied on $N(S_+({\bm w})) = N(S_-({\bm w}))$, we have:
\begin{equation}
 |S_+ ({\bm w})| \geq | N(S_+ ({\bm w})| / d_u \geq (d_l / d_u) \delta |S_-({\bm w})|
 = \rho \delta |S_-({\bm w})| \eqs.
\end{equation}
As $S_- ({\bm w})$ and $S_+ ({\bm w})$ are disjoint, we have $ |S_+ ({\bm w}) \cup |S_- ({\bm w})| \geq (1 + \rho \delta) |S_- ({\bm w})|$. As such, we choose an arbitrary subset $\tilde{S} \subseteq S_+ ({\bm w}) \cup S_- ({\bm w})$ such that $|\tilde{S}| = (1+ \rho \delta) |S_- ({\bm w})|$. Notice that  $|\tilde{S}| = (1 + \rho \delta ) |S_-({\bm w}) | \leq \alpha n$. Using the expander property again gives:
\begin{equation}
|N(\tilde{S}) | \geq \delta |E(\tilde{S},B)| \geq d_l \delta (1+ \rho \delta) |S_- ({\bm w})| > d_u | S_-({\bm w})|
\end{equation}
As $\rho \delta > (\sqrt{5}-1)/2$, the last inequality is valid as $\rho \delta (1+ \rho \delta) > 1$. 

Finally, we reach a contradiction as 
\begin{equation}
|N(\tilde{S}) | \leq |N( S_-({\bm w}) \cup S_+({\bm w}) )| = |N(S_- ({\bf w}) | \leq d_u |S_-({\bm w})|,
\end{equation}
leading to $d_u |S_-({\bm w})| > d_u |S_- ({\bm w})|$. 
The lemma is thus proven. \hfill \textbf{Q.E.D.}

Identifying that $G_{bi}$ satisfies the conditions in Lemma~\ref{lem:single}, our desirable results can be obtained. 

\section{Derivation of Problem \eqref{eq:sense_r}} \label{sec:derive}
From
\eqref{eq:obs_model}, for each $i \in [n]$, we can model the vector ${\bm y}_i$ as
\beq
{\bm y}_i = {\bm H}_i {\bm a}_i + \bm{\Delta}_i {\bm a}_i + \bm{\epsilon} \eqs,
\eeq
where $\bm{\Delta}_i$ models the error in the observed measurement matrix ${\bm H}_i$.
Let $r > 0$, 
the uncertainty set for $\bm{\Delta}_i$ is defined such that each row vector in the matrix
has a bounded norm, \ie
\beq
{\cal U}_r = \{ \bm{\Delta}_i ~:~\| {\bm d}_k \|_2 \leq r,~\forall~k \in [K+1] \} \eqs,
\eeq
where ${\bm d}_k$ is the $k$th row vector of $\bm{\Delta}_i$. 

To recover ${\bm a}_i$, we 
consider minimizing the following robust objective:
\beq 
J( \hat{\bm a}_i ) = \lambda \| \hat{\bm a}_i \|_1 + \max_{ \bm{\Delta}_i \in {\cal U}_r } \| {\bm y}_i - {\bm H}_i \hat{\bm a}_i - \bm{\Delta}_i \hat{\bm a}_i  \|_2 \eqs,
\eeq
which can be upper bounded by:
\beq \begin{split}
J( \hat{\bm a}_i ) & \leq \lambda \| \hat{\bm a}_i \|_1 + \| {\bm y}_i - {\bm H}_i \hat{\bm a}_i \|_2 + \max_{ \bm{\Delta}_i \in {\cal U}_r }  \| \bm{\Delta}_i \hat{\bm a}_i  \|_2 \\
& = \lambda \| \hat{\bm a}_i \|_1 + \| {\bm y}_i - {\bm H}_i \hat{\bm a}_i \|_2 + r \sqrt{K+1} \cdot \| \hat{\bm a}_i\|_2 \eqs,
\end{split}
\eeq
where the last equality is achieved by applying Cauchy-Schwarz and setting each row of $\bm{\Delta}_i$ to 
$r \cdot \hat{\bm a}_i / \| \hat{\bm a}_i \|_2$. 
Setting $\gamma = r \sqrt{K+1}$ and minimizing the upper bound function yields the robust
network recovery problem \eqref{eq:sense_r}.

\section{Details of experiments on empirical data} \label{sec:dream5}
This section describes the pre/post-processing procedures for the empirical support data. 

\noindent \textbf{Subspace projection for `denoising'}. 
We apply a subspace projection method as a pre-processing stage in the RIDS method. 
In particular, let the set of chips (or vectors of expression levels) taken under the 
\emph{no perturbation} condition be ${\cal C}_{nopert}$, 
we can write the gene expression levels from the $c$th chip as
\beq
\olx_{obs,c} = \olx + {\bm w}_c,~\forall~c \in {\cal C}_{nopert} \Longrightarrow  \underbrace{\big( \ldots~\olx_{obs,c}~\ldots \big)}_{\eqdef \overline{\bm X}_{obs} } = \olx {\bf 1}^\top + {\bm W}_{obs} \eqs,
\eeq 
where $\olx$ is the unperturbed steady state satisfying $\olx = {\bm A} {\bm h} (\olx)$ 
(cf.~Eq.~\eqref{eq:steady}) and ${\bm w}_c$ is modeled as an additive noise. 
When the noise is small, 
we observe that the $n \times |{\cal C}_{nopert}|$ matrix $\overline{\bm X}_{obs}$ formed 
by stacking up $\olx_{obs,c}$ horizontally is close to rank-one.
In light of this, we recover $\olx$ by taking 
the top left-singular vector of $\overline{\bm X}_{obs}$, \ie
\beq
\tilde{\olx} = \sigma_1 ( \overline{\bm X}_{obs} ) \cdot {\bm u}_1~~~\text{where}~~~\overline{\bm X}_{obs} = {\bm U} \bm{\Sigma} {\bm V}^\top \eqs,
\eeq  
and $\sigma_1 ( \overline{\bm X}_{obs} )$ is the largest singular value of $\overline{\bm X}_{obs}$. 
Notice that $\hat{\olx}$ lies in the null space of the sub-matrix $[{\bm U}]_{:, \text{2:end}}$. For each
$k \in [K]$, we
can estimate the difference ${\bm x}[k] - \olx$ by
\beq
\widetilde{ (\bm{x}[k] - \olx) } = \big( [{\bm U}]_{:, \text{2:end}}  [{\bm U}]_{:, \text{2:end}}^\top \big) \cdot \frac{1}{ | {\cal C}_{per}[k] | } \sum_{c \in {\cal C}_{per} [k] } \bm{x}_{obs,c} \eqs,
\eeq
where $\bm{x}_{obs,c}$ is the gene expression levels from chip $c$ and ${\cal C}_{per}[k]$ is the 
set of chips that corresponds to the $k$th perturbation experiment. 
Naturally, we have $\tilde{\bm x}[k] = (\widetilde{ \bm{x}[k] - \olx }) + \tilde{\olx}$. 

\noindent \textbf{Normalizing the gene expression data}. 
The vectors of gene expression level $\tilde{\olx}$ and $\tilde{\bm x}[k]$ obtained from
from the above are normalized
through dividing the vectors by the constant
\beq
 c_{norm} \eqdef \frac{1}{2} \max \Big\{ \max_{i \in [n]} ~( \tilde{\overline{x}}_i ), ~
\max_{k \in [K]} \max_{i \in [n]} ~(\tilde{x}_i [k] ) \Big\} \eqs 
\eeq
such that the values of the normalized gene expression level ranges in $[0,2]$.

\noindent \textbf{Parameters of RIDS}. For the zeros index set ${\cal S}$ in \eqref{eq:mask}, we set $\delta = 0.005$.
 The regularization parameters $\rho$ and $\gamma$ 
are set to $3 \times 10^{0}$ and $5 \times 10^{0}$ (\resp $3 \times 10^{-1}$ and $5 \times 10^{-1}$) in \eqref{eq:sense_r} for \emph{E. coli} (\resp \emph{S. cerevisiae}) network. 
Meanwhile, the AO procedure \eqref{eq:ao} aims at tackling 
\eqref{eq:sense_r} with the model response function template \eqref{eq:template}, which has several parameters to be initialized ---
for each $i \in [n]$, we initialize the AO procedure by setting 
$b_1 = 0.5$ and $b_2 = 0.5$ and the step size $\epsilon$ is set to $1 \times 10^{-2}$. 
Lastly, the AO procedure terminates after $L=10$ iterations in the numerical experiment
for a better trade-off between complexity and accuracy.

\noindent \textbf{Post-processing of the estimated GRN}. The proposed algorithm 
may recover also the negative edge weights in $\hat{\bm A}$. However, we shall
treat these negative edge weights as positive ones for the ease of benchmarking 
using the \texttt{DREAM5} evaluation script. In particular, we consider $| \hat{\bm A} |$ 
as our estimate of the GRN. 
Furthermore, we normalize the elements in $| \hat{\bm A} |$ by $\max_{ i,j } | [ \hat{\bm A} ]_{ij} |$ 
such that they range in $[0,1]$. 
For the AUROC/AUPR evaluation, we only count the top 100,000 (and 500,000) predicted links made
by our algorithm.

\section{Additional Results on Empirical Data}\label{app:add}

We present additional experiment results for the \emph{in vivo} dataset from DREAM5. 

\begin{figure}[h]
\centering
\includegraphics[width=.8\linewidth]{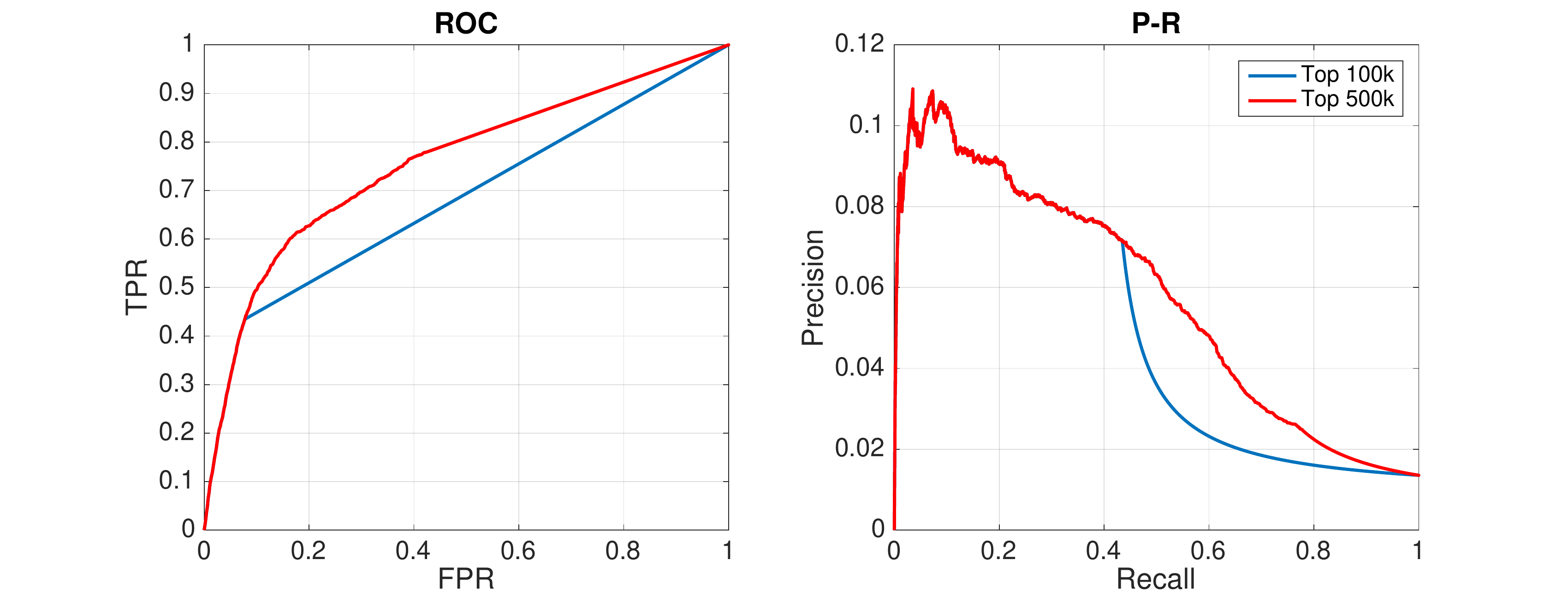}
\includegraphics[width=.8\linewidth]{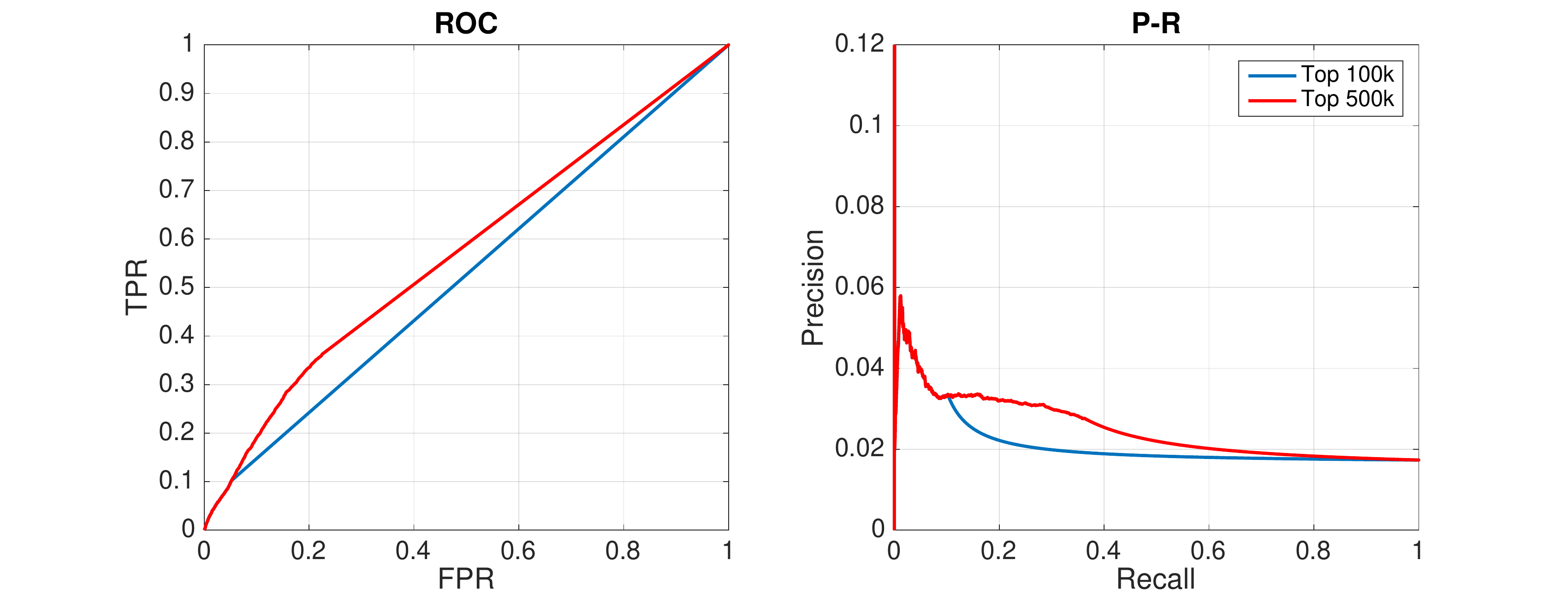}
\caption{\small
ROC and Precision-recall curves for the \emph{in vivo} networks. (Top) \emph{E. coli} network. 
(Bottom) \emph{S. cerevisiae} network. These curves are produced using the DREAM5 evaluation
script and the AUROC/AUPR values can be found in Table~\ref{tab:result}.} \label{fig:roc}
\end{figure}

\subsection{GRN recovery without the aid of transcription factors} 
For simplicity, we shall only focus on the \emph{E. coli} network.
We consider applying the
RIDS method without the help of the information provided from the list of transcription factors.
In particular, we solve \eqref{eq:sense_r} by fixing the model parameter at ${\bm b} = (0.047, 0.5893)$
and set $\rho=3$, $\gamma=5$. 

\noindent \textbf{AUROC and AUPR scores}. 
Even without any aid from the list of transcription factors, the RIDS method continues 
to yield a top AUROC/AUPR performance. In particular, 
the top $100,000$ predicted links from the identified $\hat{\bm A}$, restricted to the DREAM5
identified transcription factors,
yields an AUROC of $0.6745$ and AUPR of $0.0540$ for the \emph{E. coli} network,
as seen in Table~\ref{tab:result}. 

\begin{figure}[!t]
\centering
\includegraphics[width=.8\linewidth]{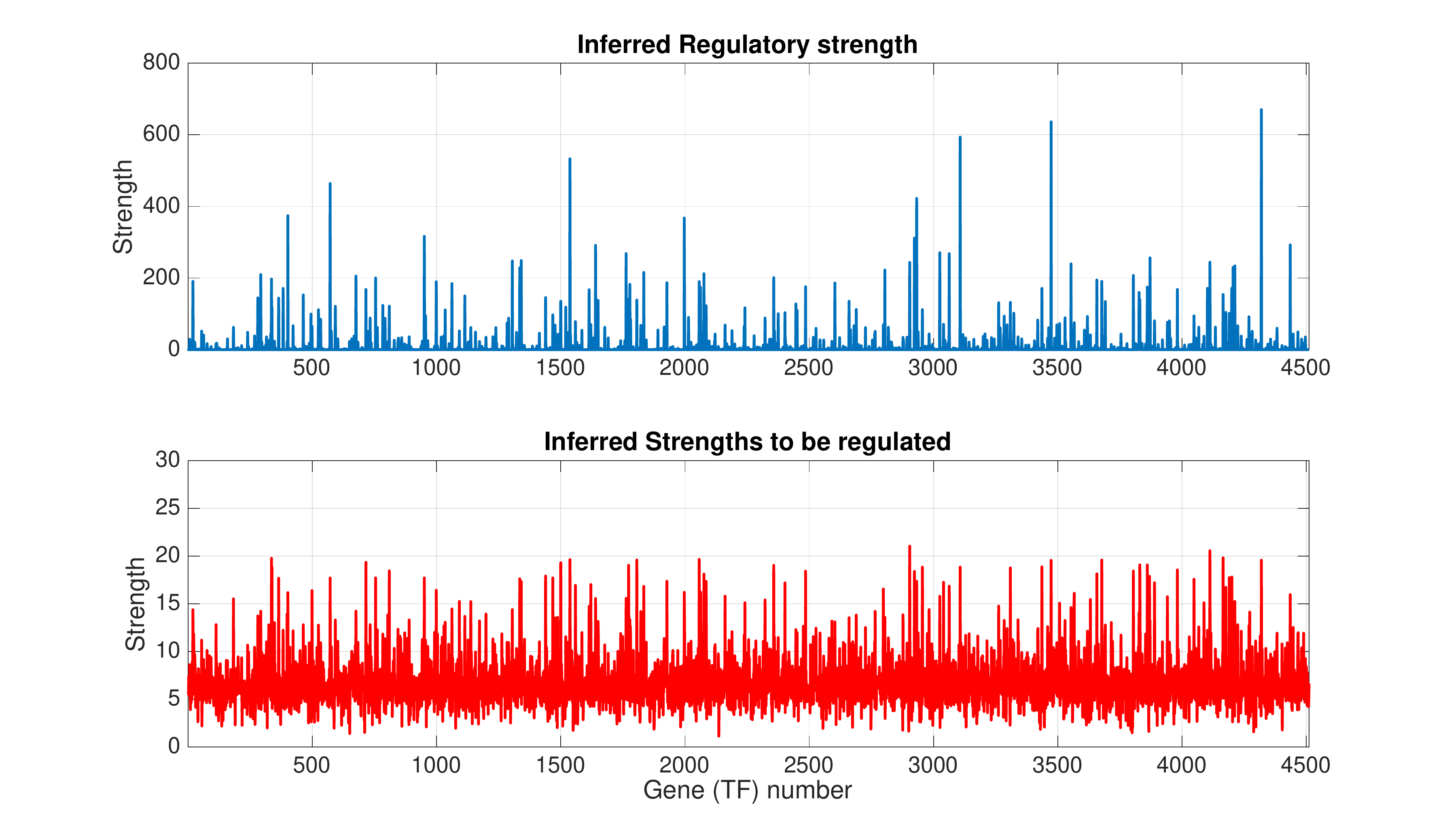}\vspace{0cm}
\caption{\small Inferring directionality of the GRN using RIDS --- comparison of 
the inferred regulatory strengths and the received regulatory strengths for each gene in \emph{E. coli}.
The top figure shows the regulatory strengths for each gene exerted on the other genes; 
the bottom figure shows the regulatory strengths \emph{received} for each gene from the
other genes.} \label{fig:strength}
\end{figure}

\noindent \textbf{Directionality of the GRN}. We test  
whether the directionality in GRNs can be inferred under the scenario described in this 
subsection \ie without using the provided list of transcription factors (TFs). 
To emphasize on the links with higher weights, 
we normalize the inferred GRN $\hat{\bm A}$ as 
$[\widetilde{\bm A}]_{ij} \eqdef | \hat{A}_{ij} |^q / \sum_{i',j'} |\hat{A}_{i'j'}|^q$ for $q =2$. 
Then we compare the following for each gene $i \in [n]$ ---
\begin{itemize}
\item Regulatory strengths --- absolute sum of the columns of $\widetilde{\bm A}$, \ie $\sum_{j=1}^n | \widetilde{A}_{ji} |$.
\item Strengths to be regulated ---  absolute sum of the rows of $\widetilde{\bm A}$, \ie $\sum_{j=1}^n | \widetilde{A}_{ij} |$.
\end{itemize} 
The comparison can be found in Figure~\ref{fig:strength}. We observe that the inferred GRN is 
directional as fewer genes have significantly higher regulatory strengths exerted on the others, 
while the regulatory strengths received for the genes are more uniformly distributed.


\end{document}